\documentclass[pra,reprint,floatfix,longbibliography,groupedaddress]{revtex4-1} 
\usepackage{amsmath}  
\usepackage{amsfonts} 
\usepackage{graphicx} 
\usepackage{bm}
\usepackage[usenames,dvipsnames]{color}
\usepackage{verbatim}
\usepackage[T1]{fontenc}
\usepackage{MnSymbol}
\usepackage{wasysym}
\usepackage{soul}
\usepackage{anyfontsize}
\usepackage{subfig} 
\usepackage{ upgreek }

\def\be{\begin{eqnarray}}
\def\ee{\end{eqnarray}}

\def\He3{$^3$He}

\def\He3{$^3$He}
\RequirePackage{lineno}

\begin{document} 

\setpagewiselinenumbers

\modulolinenumbers[5]

\title{Dual-Species Synchronous Spin-Exchange Optical Pumping}
\def\wisc{Department of Physics, University of Wisconsin-Madison, Madison, Wisconsin 53706, USA}
\author{D. A. Thrasher}

\author{ S. S. Sorensen}

\author{ T. G. Walker}
\email{tgwalker@wisc.edu} 

\affiliation{\wisc}

\date{\today}

\begin{abstract}{
 
We demonstrate a novel quantum sensor for measuring non-magnetic spin-dependent interactions. This sensor utilizes $^{131}$Xe, $^{129}$Xe, and $^{85}$Rb which are continuously polarized transverse to a pulsed bias field. The transverse geometry of this spin-exchange pumped comagnetometer suppresses longitudinal polarization, which is an important source of systematic error. Simultaneous excitation of both Xe isotopes is accomplished by frequency modulating the repetition rate of the bias field pulses at subharmonics of the Xe Larmor resonance frequencies. The area of each bias pulse causes $2\pi$ Larmor precession of the Rb. We present continuous dual-species Xe excitation and discuss a temperature-dependent wall interaction that limits the $^{129}$Xe polarization. The Rb atoms serve as an embedded magnetometer for detection of the Xe precession. We discuss Rb magnetometer phase shifts, and show that even first-order treatments of these phase shifts can result in order-of-magnitude improvements in the achieved field suppression when performing comagnetometry. The sensing bandwidth of the presented device is 1 Hz, and we demonstrate a white-noise level of 7 $\mu$Hz/$\sqrt{\text{Hz}}$ and a bias instability of $\sim1$ $\mu$Hz. 
}\end{abstract}

\maketitle



\section{Introduction} \label{Introduction}

Spin-exchange (SE) pumped comagnetometers~\cite{Limes2018,Walker2016} enable precise sensing of spin-dependent phenomena with benchtop-scale experiments. Such devices have found broad use in the precision measurement community, with applications ranging from measuring inertial rotation~\cite{Walker2016,Kornack2005,Jiang2018,Karwacki1980} to placing upper bounds on spin-mass couplings~\cite{Bulatowicz2013,Lee2018}, Lorentz violations~\cite{Allmendinger2014,Romalis2014,Smiciklas2011,Brown2010,Glenday2008}, and atomic electric dipole moments~\cite{Rosenberry2001,Allmendinger2019,Sachdeva2019}. 

The potential for miniaturization has made gyroscope implementations a subject of interest in industrial research~\cite{Walker2016,Volk80b,Kwon1981}. Indeed, advances in chip-scale technologies have opened the door for SE pumped comagnetometers whose performance is predicted to scale favorably with the size of the sensor compared to alternative devices such as ring laser gyroscopes~\cite{Donley2010}. Practical inertial navigation devices must provide continuous readout. The device presented in this work was designed to suppress systematic errors from longitudinal polarization while satisfying the constraint of continuous operation.

SE pumped comagnetometers consist of co-located ensembles of noble gas nuclei and alkali-metal atoms which are spin polarized in the presence of a magnetic field~\cite{Walker1997}. Suppose an ensemble of two spin-exchange optically pumped (SEOP) noble gas species ($a$ and $b$) are each subject to a common magnetic field $B_{z}$ and some spin-dependent interaction $X$. The Larmor resonance frequency of each isotope obeys~\cite{Walker2016,Limes2019,Terrano2019,Petrov2019}
\begin{subequations}
\label{OmegaXe} 
\begin{eqnarray}
\Omega^{a} = \gamma^{a}( B_{z}+b_S^a S_z+ b_b^a K_z^b) + X_z^{a},
\\
\Omega^{b} = \gamma^{b}( B_{z}+b_S^b S_z+ b_a^b K_z^a) + X_z^{b},
\end{eqnarray}
\end{subequations}
where $\gamma$ is the gyromagnetic ratio, $S$ and $K$ are the respective alkali-metal and noble gas polarizations, $z$ subscripts refer to the longitudinal components (\textit{i.e.}, parallel to the bias field direction), and $b^i_j$ is the SE coefficient~\cite{Schaefer89} characterizing the influence of $j$'s polarization on $i$. Given a known $\rho = \gamma^a$/$\gamma^b$, $\Omega^a$ and $\Omega^b$ can be simultaneously measured in order to suppress $B_{z}$ while retaining sensitivity to $X_z^a$ and $X_z^b$~\cite{Chupp1988}. The longitudinal polarizations $S_z$ and $K_z$ are important sources of systematic error in SE pumped comagnetometers~\cite{Bulatowicz2013,Limes2019,Terrano2019}.
 
The embedded alkali-metal atoms can be used for quantum non-demolition readout~\cite{Takahashi1999,Katz2019} of the noble gas precession. During atomic collisions, a Fermi-contact interaction enhances the field experienced by the alkali-metal atoms due to the the polarized noble gas nuclei~\cite{Nahlawi2019}. This enhancement factor directly improves the signal-to-noise-ratio (SNR) as classical fields are not similarly enhanced. Using the alkali-metal atoms also enables miniaturization by eliminating the need for an exterior pick-up coil (such as a SQUID~\cite{Allmendinger2019,Sachdeva2019}). 

The device demonstrated in this work is a SE pumped $^{131}$Xe-$^{129}$Xe comagnetometer which produces no first-order time-averaged $S_z$ or $K_z$ such that the comagnetometer signal
\be
\xi \equiv  {\rho \Omega^b - \Omega^a\over 1+\rho}  \approx {\rho X_z^b - X_z^a\over 1+\rho}
\ee
is independent of magnetic fields, where the superscripts $a$ and $b$ refer to $^{129}$Xe and $^{131}$Xe, respectively, and $\rho = 3.373417(38)$~\cite{Makulski2015}. In the context of rotation sensing, we set $X_z^{b} = \omega^R$ and $X_z^{a} = -\omega^R$~\cite{Walker2016} such that $\xi$ becomes the rotation frequency $\omega^R$. The noble gas nuclei are continuously polarized perpendicular to a frequency-modulated pulsed bias field, utilizing a dual-species extension of synchronous SEOP~\cite{Korver2015,Thrasher2019}. We demonstrate that the $^{131}$Xe and $^{129}$Xe Larmor precession frequencies are sufficiently correlated to resolve a white frequency noise level of 7 $\mu$Hz /$\sqrt{\text{Hz}}$ and achieve a low-frequency field noise suppression in excess of $10^3$. The scale factor between the Larmor resonance frequencies and the spin-dependent interactions is determined predominantly by fundamental physical constants, namely the gyromagnetic ratios. It is, to a good approximation~\cite{Brinkmann62}, independent of the details of our apparatus (such as temperature, gas pressures, etc.). This means that comagnetometry can be performed without the need for calibration.

\begin{figure}
\includegraphics[width = 3.2in]{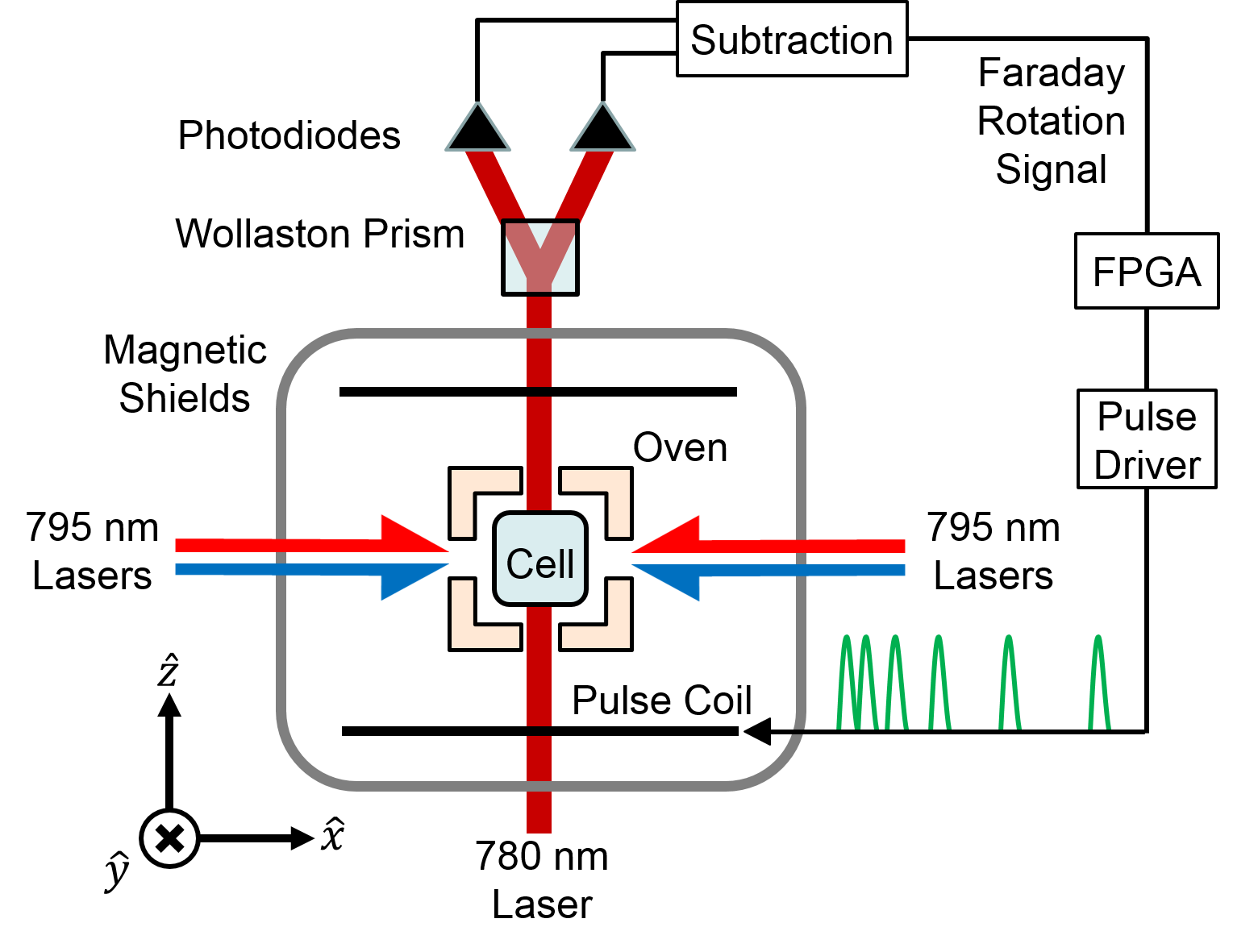}
\caption{Schematic of apparatus. Field shim coils are not shown. The green trace depicts the frequency modulated bias field pulses. }\label{schem}
\end{figure}

Figure~\ref{schem} shows a schematic of the experimental apparatus. We optically pump $^{85}$Rb atoms along $\hat{x}$, and noble gas nuclei are polarized via SE collisions with the polarized Rb atoms. The bias field is oriented along $\hat{z}$ (transverse to the Rb polarization) and is applied as a series of low-duty-cycle pulses. The pulse area is set such that the Rb spins precess $2\pi$ radians during each bias pulse. The Xe isotopes precess only $\sim 2\pi/10^3$ radians per pulse (owing to their much smaller magnetic moments). As such, we can approximate the effective bias field experienced by the Xe isotopes due to the pulses as a continuous function,
\begin{equation}
B_{p}(t) =\omega_{p}(t) / \gamma^S,
\end{equation}
where $\omega_{p}(t)$ is the repetition rate of the $2 \pi$ pulses and $\hbar \gamma^S=2\mu_B/(2I+1)$, where $I=5/2$ is the $^{85}$Rb nuclear spin and $\mu_B$ is the Bohr magneton. The repetition rate of the bias field pulses is modulated at linear combinations of $\Omega^a$ and $\Omega^b$ in order to simultaneously excite the magnetic resonance of both noble gas species.

Continuous detection of the noble gas precession is accomplished using the embedded Rb magnetometer, which is effectively at zero-field due to the bias field being applied as low-duty-cycle $2\pi$ pulses~\cite{Korver2013}. We monitor $S_z\propto K_y$ by measuring the Faraday rotation of a linearly polarized probe laser which propagates along $\hat{z}$ (parallel to the bias field). We demodulate the $S_z$ signal digitally using an FPGA. Resonance for each Xe species is maintained by adjusting the modulation frequencies of the pulsed bias field repetition rate, and $\xi$ is computed using the Xe resonance frequencies determined by this feedback.

This paper expands on the work presented in Ref.~\cite{Thrasher2019} and proceeds as follows. Section~\ref{dualSEOP} describes the simultaneous excitation, detection, and demodulation of $^{131}$Xe and $^{129}$Xe precession. We give details of the experimental apparatus and demonstrate dual-species Xe excitation. In Sec.~\ref{comag}, we perform comagnetometry and present the field suppression and stability of the system. Section~\ref{outlook} provides an outlook for possible future studies, and we conclude in Sec.~\ref{conclusion} with a summary of the presented work.

\section{Excitation and Detection}  \label{dualSEOP}

\subsection{Principles}


We assume purely transverse optical pumping of the Rb along $\hat x$. We also assume that transverse fields experienced by the Xe, including the SE field $b_S^K S_x$, are well-nulled, such that the longitudinal components of the Xe polarizations are negligible~\cite{Korver2015}.

The transverse components $K_+=K_x+iK_y$ of the nuclear spin polarization for each noble gas species obey
\be\label{BE1}
{dK_+\over dt}=-(\mp i\Omega+\Gamma_2)K_+ +\Gamma_{S}^K S_+
\ee
where $\Omega=\gamma B_z+ X_z$ is the Larmor resonance frequency, arising from both magnetic field $B_z$ and spin-dependent phenomena $X_z$. The SE rate constant is $\Gamma_{S}^K$, and the transverse relaxation rate is $\Gamma_2$. The precession direction is encoded in the sign in front of $\Omega$ (top is $^{129}$Xe, bottom is $^{131}$Xe). The frequency shift of the $^{131}$Xe due to quadrupole interactions is assumed to be independent of $B_z$. Such quadrupole effects are included in $X_z^b$.

Simultaneous excitation of multiple noble gas species via SE collisions with the Rb can be accomplished by modulating either $S_+$ or $B_z$. Previously we demonstrated single-species excitation by modulating $S_+$~\cite{Korver2015}. While an extension of that approach is possible, here we perform dual-species excitation by modulating $B_z$. In this work, we modulate $B_z$ by modulating the pulse repetition rate $\omega_{p}$. This is advantageous because changes in $\omega_p$ impact the Rb magnetometer gain far less than do changes in DC fields.

The $\hat{z}$ component of the magnetic field is $B_z=B_{z0}+B_p(t)$, which includes the stray field $B_{z0}$ and the bias field from the $2\pi$ pulses $B_p(t) = B_{p0}+B_m(t)$, where $B_m$ is the modulated part of the pulsed field. The Xe nuclei experience both $B_{z0}$ and $B_p$, but because the Rb atoms precess by $2\pi$ during each bias pulse, the Rb atoms experience predominantly $B_{z0}$.

We find that the Rb magnetometer gain depends on the pulse repetition rate. We attribute this to changes in the effective relaxation rate of the magnetometer as the pulse repetition rate (and hence duty cycle) is varied~\cite{Korver2013}. Though the bias field is zero most of the time, any SE collisions that occur during a bias pulse will contribute to relaxation. Hence, the magnetometer gain will be increasingly degraded as pulses are applied more frequently. For the $\omega_p$ modulation presented in this work, the observed magnetometer gain varied up to a factor of two.

\begin{figure}
\includegraphics[width=3.5in]{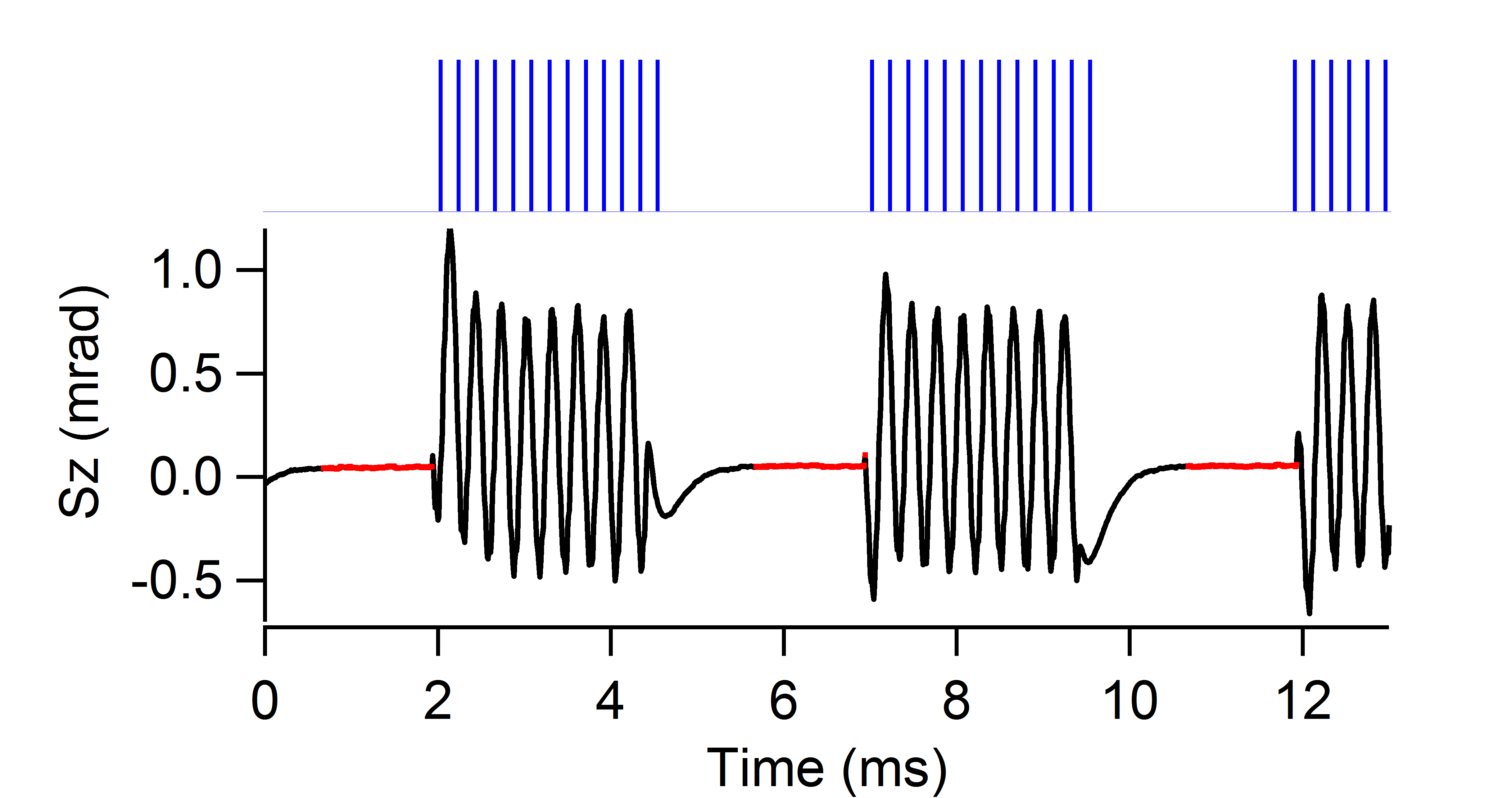}
\caption{Influence of bias pulse gating on Rb magnetometry. (Top) Time dependence of the few percent duty cycle bias pulses. (Bottom) Measured time dependence of raw $S_z$ signal from polarimeter. Red data are averaged together to measure the Xe precession at an effective sampling rate of $\omega_3$. The $\approx 3$kHz oscillation present when pulsing comes from interference between the AC heater drive and the pulses. }\label{gate}
\end{figure}

\begin{figure}
\includegraphics[width = 3.5in]{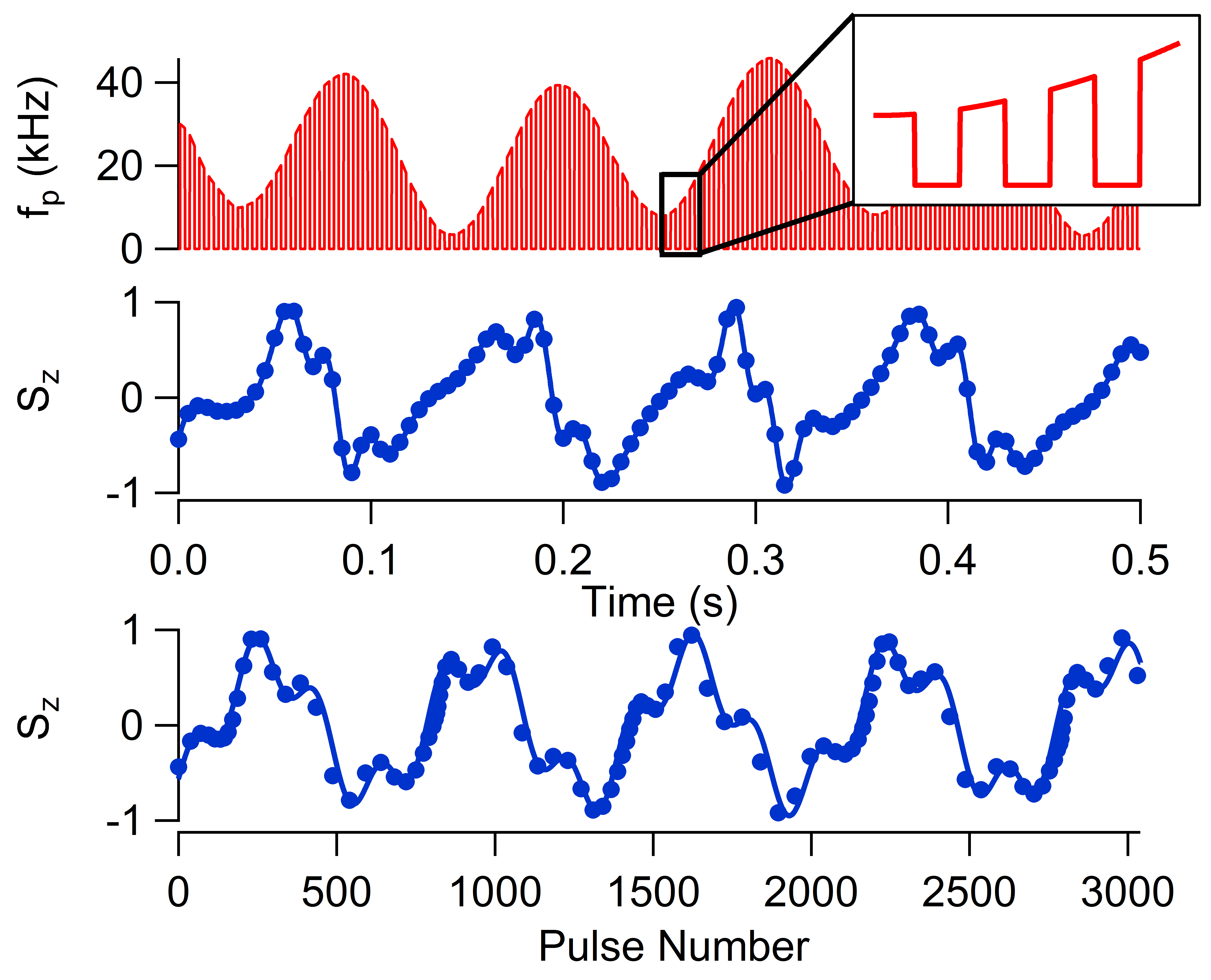}
\caption{Time dependence of bias pulse repetition rate (top) and normalized $S_z$ sampled at $\omega_3$ (middle) for $\Delta^a = \Delta^b = 0$. Corresponding pulse number dependence of normalized $S_z$ (bottom). Filled circles are measured data. Lines are theory fits to the data. Inset depicts gating of the bias pulses.}\label{sigs}
\end{figure}

We suppress the influence of gain modulation on the detection by gating the 2$\pi$ pulses. After the pulses are gated off, we wait for the magnetometer gain to recover before recording the Faraday signal (see Fig.~\ref{gate}). If we do not gate the pulses, we see a background on the recorded Faraday signal which corresponds to harmonics of the pulsing modulation frequencies. We find that waiting twice the $1/e$ recovery time of the Rb magnetometer before recording the Faraday signal suppresses this background. We modulate the $2\pi$ pulse repetition rate (depicted in Fig.~\ref{sigs}) as
\be \label{wpeq}
\omega_p(t) = \omega_{p0}g(t)(1+ b_1 \cos(\omega_1t)+b_2 \cos(\omega_2 t)),
\ee
where $g(t) = (\rm{sign}(\cos(\omega_3 t))+1)$ is the time dependence of the gating, $\omega_1 = \omega_d^{b}$ and $\omega_2 = \omega_d^{a} - 3\omega_d^{b}$ determine the Xe drive frequencies, and $b_1$ and $b_2$ set the depth of modulation.

\subsubsection{Influence of Rb on drive, detection}
The Rb atoms serve two primary purposes in our system: we use the DC transversely polarized Rb $S_+$ to polarize the Xe nuclei and the AC longitudinally polarized Rb $S_z$ to detect the Xe precession. For $\Omega<< \Gamma'$ the time-average solution to the Bloch equations for the Rb polarization
\be \label{BE2}
\mathbf{S} = {\mathbf{R}\Gamma'+\mathbf{\Omega} \times \mathbf{R}+\mathbf{\Omega}(\mathbf{\Omega}\cdot \mathbf{R})/\Gamma' \over \Gamma'^2+\mathbf{\Omega}^2}
\ee
can be expanded as
\be \label{Seq}
\mathbf{S} = {\mathbf{R}\over \Gamma'} +{\mathbf{\Omega} \times \mathbf{R} \over\Gamma'^2}+{\mathbf{\Omega} \times \mathbf{\Omega} \times \mathbf{R}\over\Gamma'^3}+..., 
\ee 
where $\mathbf{R}$ is the pumping rate, $\Gamma'$ is the total relaxation rate (including pumping), and $\mathbf{\Omega} = \gamma^S \mathbf{B}$ where $\mathbf{B}$ is the magnetic field experienced by the Rb. This assumes negligible back polarization from the Xe to the Rb~\cite{Bhaskar80b,Limes2018}, and that $\mathbf{K}$ precesses slowly such that $S_z$ responds adiabatically. Since we optically pump along $\hat{x}$ we have $\mathbf{R}=R\hat{x}$, and

\begin{subequations}
\label{allequations} 
\begin{eqnarray}
S_+ = {R\over\Gamma'}+i{R \Omega_z\over \Gamma'^2} = {R \over \Gamma'}e^{i\epsilon_z},
\\
S_z = -{R\over \Gamma'^2}(\Omega_y -{\Omega_z\over \Gamma'} \Omega_x)={-R\over \Gamma'^2 }\text{Im}[\gamma^S b_K^S K_+e^{-i\epsilon_z}],
\end{eqnarray}
\end{subequations}
where 
\be \label{ezdef}
\epsilon_z = \tan^{-1}({ S_y \over S_x}) \equiv \tan^{-1}({ B_{z0} \over B_w})<<1
\ee
is the magnetometer phase shift. $B_w=\Gamma'$/$\gamma^S$ is the magnetic width of the magnetometer. We see that though the Rb atoms are optically pumped along $\hat{x}$, the field $B_{z0}$ causes the Rb polarization to rotate slightly into $\hat{y}$ so that the transverse spin is phase shifted. We will show that this phase shift appears on the detected phase of rotating transverse fields (see Eq.~\ref{demodph}) and must be accounted for in order to achieve optimal field suppression in the comagnetometer signal (see Sec.~\ref{FSF}). 

\subsubsection{Magnitude and phase of $K_+$}
We write the nuclear spin polarization using a phasor representation $K_+ = K_{\perp} e^{ \pm i\phi}$. We measure the difference $\delta = \phi - \alpha$ between the instantaneous Xe phase $\phi$ and a reference phase $\alpha = \int(\omega_d + \gamma B_m) dt$ which is the phase the Xe would have if the only fields present were the pulsing fields and if $\Delta \equiv\omega_d - \Omega_0=0$ with $\Omega_0=\gamma(B_{z0}+B_{p0})+X_z$. To first order in $\delta$ and $\epsilon_z$, the real part of Eq.~\ref{BE1} is
\be\label{KP1}
{dK_{\perp}\over dt}=-\Gamma_2 K_{\perp} +\Gamma_{S}^K S_{\perp} [\cos(\alpha)-(\delta \mp \epsilon_z)\sin(\alpha)].
\ee
To find the time averages of $\cos(\alpha)$ and $\sin(\alpha)$ (and thereby arrive at a steady-state solution for $K_{\perp}$), consider the time average of $e^{i\alpha}$ for $B_m=B_1\cos(\omega_1t)+B_2\cos(\omega_2t)$. Making substitutions using the Jacobi-Anger expansion
\be
e^{i z sin\theta}=\Sigma_{n=-\infty}^{\infty} J_n(z)e^{i n \theta},
\ee
where $ J_n(z)$ is the $n$-th Bessel function of the first kind, and keeping only terms of the sums that would mix to give a contribution at DC, we find the time averages $\overline{\cos(\alpha)}=J_{-p}({\gamma B_1\over \omega_1})J_{-q}({\gamma B_2\over \omega_2})$ and $\overline{\sin(\alpha)}=0$. The steady-state solution for $K_{\perp}$ is then
\be \label{KPss}
K_{\perp}={\Gamma_S^K S_{\perp}\over \Gamma_2}J_{-p}({\gamma B_1\over \omega_1})J_{-q}({\gamma B_2\over \omega_2})
\ee
with $p=3$, $q=1$ for $^{129}$Xe and $p=1$, $q=0$ for $^{131}$Xe chosen to satisfy the resonance condition $\omega_d=p\omega_1+q\omega_2$.

The imaginary part of Eq.~\ref{BE1} is
\be\label{dphi}
{d\phi\over dt} =\Omega-{\Gamma_S^K S_{\perp} \over K_{\perp}}\sin(\phi \mp \epsilon_z).
\ee
To get an expression for $\delta$, we make substitutions as given above and use the steady-state solution for $K_{\perp}$. To first order in $\delta$ and $\epsilon_z$, we arrive at
\be\label{delt}
{d\delta\over dt} =-\Delta-\Gamma_2(\delta \mp \epsilon_z).
\ee
The sign in front of $\epsilon_z$ is isotope dependent because the Xe isotopes precess in opposite directions.

\subsubsection{Detection and demodulation} \label{detdem}
From Eq.~\ref{allequations} and Eq.~\ref{KPss}, the magnetometer signal $S_z$ can be written as
\be\label{Szdet}
S_z = A^{a}_{\perp} \sin(\delta^{a}+\alpha^a - \epsilon_z)+A^{b}_{\perp} \sin(\delta^{b}+\alpha^b + \epsilon_z),
\ee
where $A_\perp= -\gamma^S  b^S_K K_\perp R$/$\Gamma'^2$. Note that the Rb precession phase includes both the Xe precession phase and the magnetometer phase shift. 

The precession phase of each isotope can be extracted from $S_z$ by demodulation with $\cos(\alpha)$. For example,
\begin{multline} \label{demodph}
\int{d\alpha^b S_z \cos(\alpha^b)}=\int{dt {d\alpha^b\over dt}S_z \cos(\alpha^b)}\\ = A^b_{\perp}\sin(\delta^b+\epsilon_z)+ res.
\end{multline}
We sample evenly in time for experimental convenience, necessitating the ${d\alpha\over dt}$ in the demodulation. Note the orthogonality between ${d\alpha^a\over dt} \cos(\alpha^a)$ and ${d\alpha^b\over dt} \cos(\alpha^b)$ which prevents isotope $a$'s phase information from showing up on isotope $b$'s detection channel and vice versa. The $S_z$ signal is sampled at $\omega_3$ and then low-pass filtered to prevent aliasing. The time average is approximated using a moving average over $N$ data points, where $N$ is chosen to suppress the high frequency residuals of the demodulation. 

\subsection{Apparatus}
The experimental apparatus is similar to that described previously in Ref.~\cite{Korver2015}. An 8 mm cubic Pyrex cell filled with 40 Torr enriched Xe and 50 Torr N$_2$ with a hydride coating~\cite{Kwon1981} is mounted in a ceramic housing which is secured inside a 3D-printed magnetic field coil rig, which is itself mounted in a three-layer cylindrical mu-metal shield (see Fig.~\ref{schem}). The coil rig consists of three square orthogonal Helmholtz coils (for nulling stray fields) and a multipole pulsing coil designed specifically to have low inductance, high field uniformity across the cell, and low magnetic moment (to reduce coupling to the shield endcaps). The multipole pulsing coil consists of two pairs of square coils wound in series with opposite polarity. The circuit used to drive the pulsing coil was custom-made using a modified H-bridge design. The bias field requires short pulses ($<5$ $\mu$sec) of $\approx$ 1 Ampere peak current. To facilitate modulated-$\omega_p$ operation, the circuit was designed such that the Rb Larmor precession produced by each pulse is largely independent of the pulse repetition rate. The cell is heated to $\sim 120$ \textdegree C by applying AC current at $\sim 150$ kHz to four pairs of heater coils printed on each face of the cell's ceramic housing. 

To perform optical pumping of the Rb, the outputs of two distributed feedback laser diodes tuned near the Rb D1 transition (one on either side of resonance) are overlapped. The resulting beam is then split and circularly polarized so that half of each laser's power is directed to the cell from opposing directions ($\pm \hat{x}$), thereby reducing intensity gradients. The total pump light incident to the cell is nominally 30 mW. The power and detuning of each pump laser is chosen to null the average AC Stark field seen by the Rb. The average pump detuning is $\approx 12$ GHz. 

To detect $S_z$, approximately one mW of linearly polarized light from the output of a third distributed feedback laser diode, tuned near the Rb D2 line, is directed through the center of the cell and parallel to $\hat{z}$. Its polarization is analyzed via a balanced polarimeter. An FPGA synced to a commercial Rb time standard is used to synthesize the $2\pi$ pulse triggers, demodulate the digitized Faraday rotation signal, and perform feedback to correct the pulse repetition rate modulation frequencies so that the measured phase of each Xe isotope is kept equal to zero. 

\subsection{Xe signals}

\begin{figure}
\includegraphics[width = 3.5in]{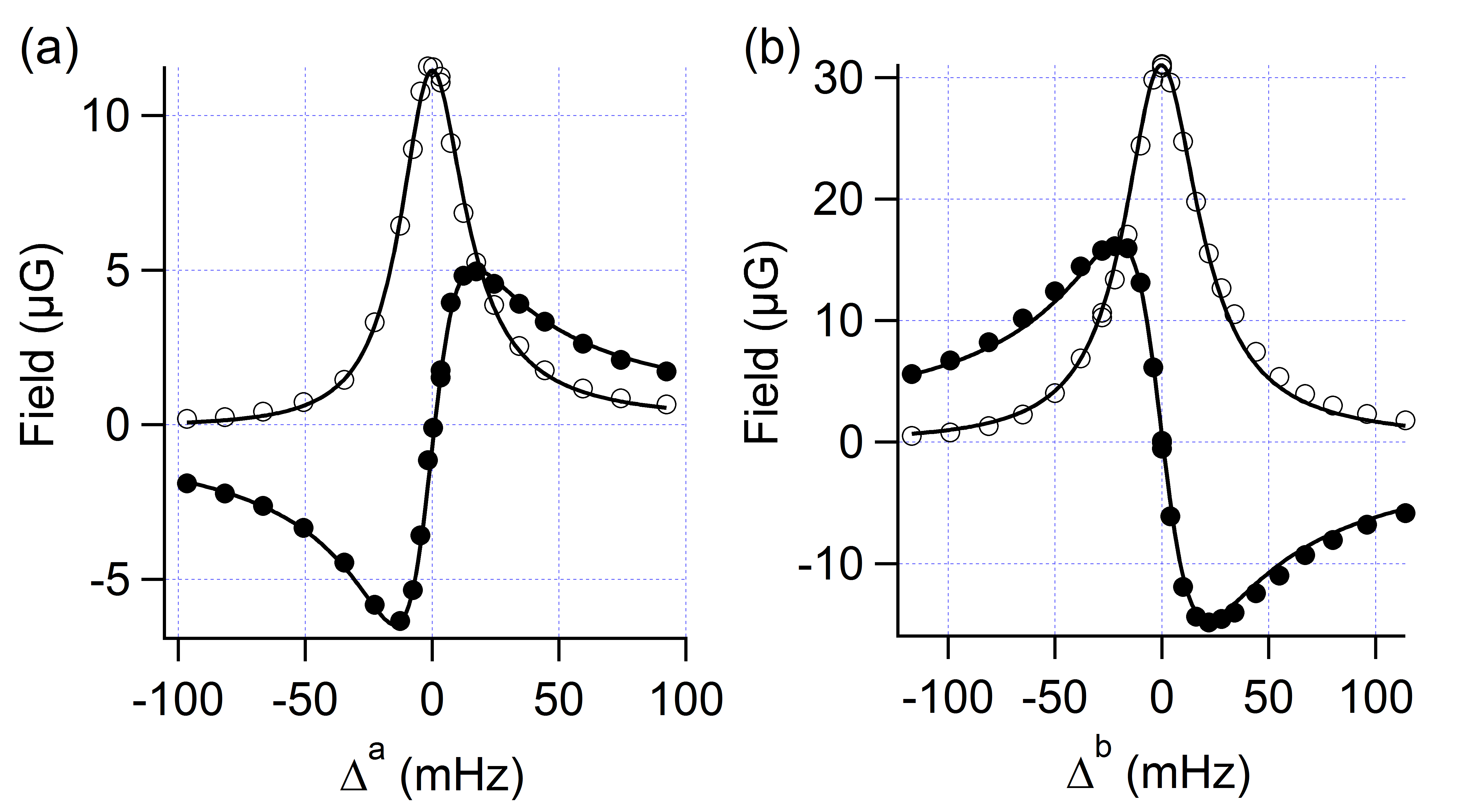}
\caption{Xe NMR Lineshapes. (a) $b_a^S K_{\perp}^a$ vs $\Delta^a$. (b) $b_b^S K_{\perp}^b$ vs $\Delta^b$.  Filled circles are measured data. Lines are Lorentzian fits to the data.}\label{nmrx}
\end{figure}

The demodulated $S_z$ vs. $\delta$ for each isotope is shown in Fig.~\ref{nmrx}.  The data were acquired with one isotope driven on resonance while the other isotope's detuning was varied. The field modulation parameters were, $\omega_3 = 2\pi\times200$ Hz, $b_1=0.73$ and $b_2=0.15$, and $\omega_{p0}\approx 2\pi\times13.2$ kHz, resulting in average precession frequencies of $\approx 33.3$ Hz and $\approx 9.9$ Hz for $^{129}$Xe and $^{131}$Xe, respectively.  Under these conditions, the amplitude of $B_m$ was approximately $10 B_w$. Only the modulation of $2\pi$ pulses allows for the Xe to experience such a large modulation while preserving the fidelity of the Rb magnetometer. The modulation parameters $b_1$ and $b_2$ were chosen to optimize the simultaneous excitation of both Xe species. Limiting $b_1+b_2<1$ avoids producing Xe precessions near the $1/f$ detection noise without requiring reversals of the pulsing direction. For our choice of $\omega_{p0}$, $b_1+b_2=0.88$ ensures that the instantaneous precession frequency of the $^{131}$Xe never goes below 1 Hz. The moving average filter most strongly attenuates frequency content at integer multiples of $\omega_3/N$ (where $\omega_3$ sets the sampling rate). Since $\rho$ was within $0.05\%$ of $27/8$, we used the moving average filter to suppress residuals of the demodulation at $\omega_d^a$, $\omega_1=\omega_d^b$, and $\omega_2$ by setting $\omega_d^a=\omega_3/6$ and $N=27 \times \omega_3/\omega_d^a=162$.

For $^{131}$Xe we measure $b_b^S K^b_{\perp}\approx$ 30 $\mu$G or 0.1\% polarization and a linewidth of $21.2(3)$ mHz. For $^{129}$Xe we measure $b_a^S K^a_{\perp}\approx$ 10 $\mu$G or 0.3\% polarization and a linewidth of $15.6(3)$ mHz. From Eq.~\ref{KPss} and esitmates similar to Ref.~\cite{Walker2016}, we anticipate polarizations of 0.05\% for $^{131}$Xe and 0.2\% for $^{129}$Xe.

\begin{figure}
\includegraphics[width = 2.5in]{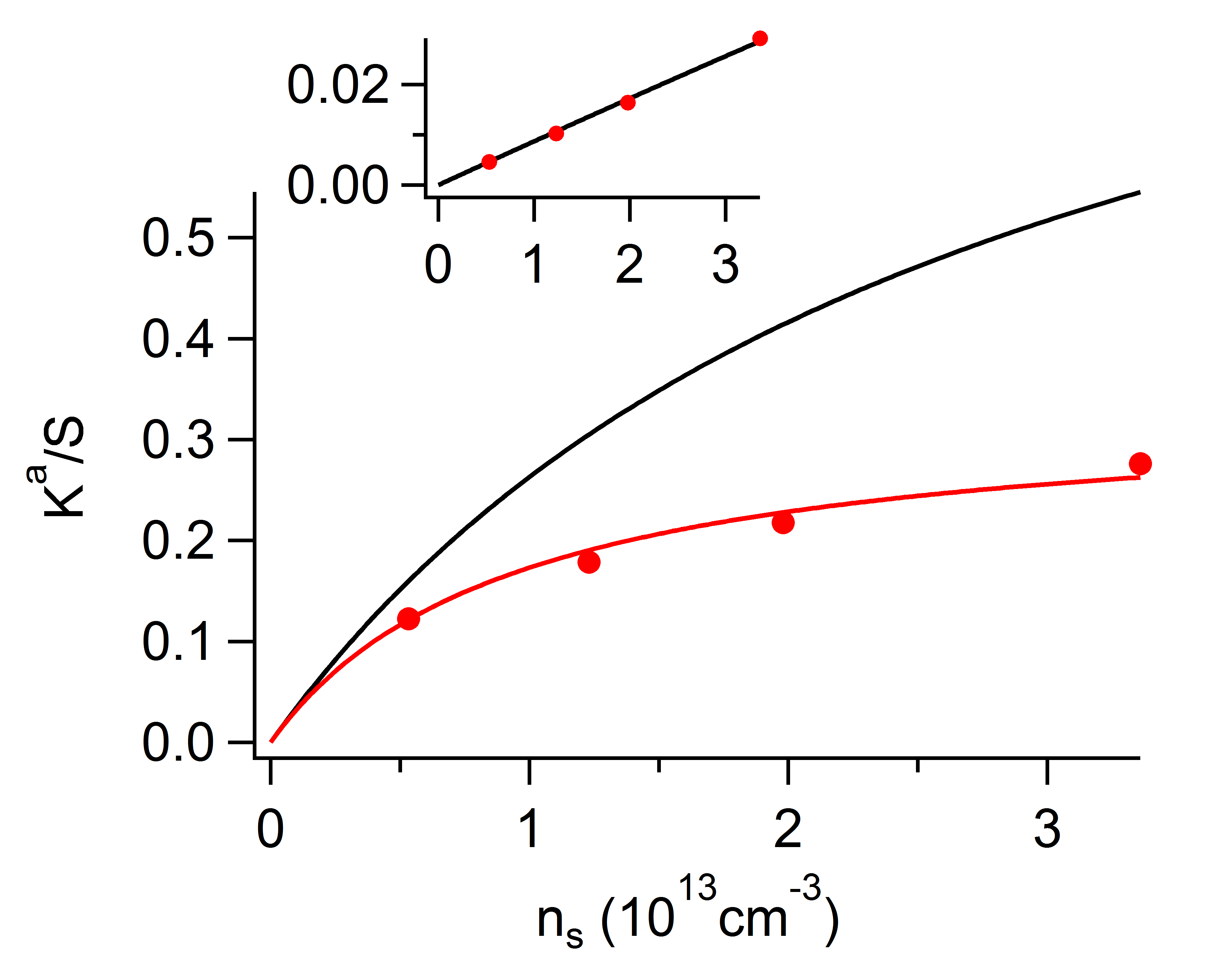}
\caption{Ratio of Xe to Rb polarization vs Rb density. Inset depicts $K^b$/$S$. Filled circles are measured data. Black lines are theoretical curves which lack temperature-dependent wall relaxation. Red line is a fit of $K^a/S$ vs Rb density which includes an anomalous wall relaxation free parameter.}\label{xfactor}
\end{figure}

The measured $^{129}$Xe polarization is substantially lower than if it were limited only by relaxation due to SE collisions. A possible explanation for this discrepency is an anomalous temperature-dependent wall relaxation mechanism similar to what has been reported in Rb-$^{3}$He cells~\cite{Babcock2006}.  We fit the Xe-Rb polarization ratio as a function of Rb density $n_S$ with
\be \label{xfactorEQ}
{K_{\perp}\over S_{\perp}}={\Gamma_{S}^K\over \Gamma_1}={\kappa_{S}^K n_S\over \kappa_{S}^K n_S (1+x)+\Gamma_{W0}},
\ee
where $\kappa_S^K$ is the SE rate coefficient, and the wall relaxation is $\Gamma_W=\Gamma_{W0}+x\Gamma_S^K$ with room-temperature relaxation $\Gamma_{W0}$. The $x$ parameterizes the temperature dependence. Figure~\ref{xfactor} shows the diminishing gains in $^{129}$Xe polarization as the Rb density is increased, corresponding to $x=2.0(1)$ in our cell. Previous measurement of a nominally spherical 8 mm cell (32 Torr $^{129}$Xe, 8 Torr $^{131}$Xe, 300 Torr N$_2$, and $^{85}$Rb) gave $x=3.4$ for $^{129}$Xe~\cite{Korver2015}. $^{131}$Xe does not exhibit such a temperature-dependent effect; its polarization scales linearly with Rb density (see inset of Fig.~\ref{xfactor}). As this is a surface effect, it is likely related to the Rb hydride coating of our vapor cell. Further study is needed to understand the exact mechanism behind $x$. Such study will require measurements of vapor cells of various compositions and geometries.

Free induction decay studies suggest the presence of $\approx 30$ mHz of first order quadrupole produced by the cell assymetry caused by the cell's pull-off stem~\cite{Wu1988}. However, the measured $^{131}$Xe signal exhibits no sign of quadrupole beating~\cite{Donley2009,Bulatowicz2013} when excited via synchronous SEOP. Furthermore, the average (over many data sets) of the ratio of Xe resonance frequencies is 3.3727. Supposing that the discrepancy between the measured ratio of gyro-magetic ratios and that reported in Ref.~\cite{Makulski2015} is due solely to quadrupole, the quadrupole shift would be about 2 mHz. We believe the suppression of first order quadrupole is generic to transverse polarization and plan to explore this interesting observation in future work. 

\section{Comagnetometry} \label{comag}


Simultaneous measurements of the Xe resonance frequencies $\Omega^a$ and $\Omega^b$ are required for the computation of the comagnetometer signal $\xi$. In our system, we measure the resonance frequencies using feedback. The phase $\delta$ of the Xe transverse polarization is much more sensitive to small fluctuations (or rotations) about the bias field than the magnitude $K_{\perp}$, and is therefore a useful error signal for feedback. Performing a Fourier transform of Eq.~\ref{delt} gives
\be \label{delt2}
\tilde{\delta} = -{\tilde{\Delta} \mp \Gamma_2 \tilde{\epsilon}_z \over i\omega +\Gamma_2}.
\ee
We stabilize the drive frequency of each isotope by servoing~\cite{Bechhoefer2005} the detected phase $\tilde{\delta} \mp \tilde{\epsilon}_z$. Doing so reduces uncertainty in the transformation from phase to frequency for each isotope. The feedback can be written
\be
\tilde{\omega}_d =(\tilde{\delta}\mp \tilde{\epsilon}_z) \tilde{G},
\ee
where $\tilde{G}$ is the frequency dependent feedback gain (units of 2$\pi$ Hz/rad). The finite gain corrected drive frequencies are then
\be \label{upomeg}
\tilde{\upomega}_d = \tilde{\omega}_d\tilde{G}^{-1}(i \omega +\Gamma_2+\tilde{G})=\tilde{\Omega}_0 \mp i\omega \tilde{\epsilon}_z.
\ee
The comagnetometer signal $\xi^{\prime}$ is computed as
\be \label{xiprime}
\tilde{\xi}^{\prime} \equiv  {\rho \tilde{\upomega}_d^{b} - \tilde{\upomega}_d^{a}\over 1+\rho}= \tilde{\xi}+i\omega \tilde{\epsilon}_z,
\ee
where $\upomega^a,\upomega^b>0$. We see that the derivative of $\epsilon_z$ appears just like a rotation would in $\xi^{\prime}$.

To extract $\xi$ from $\xi^{\prime}$, we need to know $\epsilon_z$. We assumed that $\tilde{B}_{z0}$ was the dominant contribution to $\tilde{\epsilon}_z$ and used the sum of the finite gain corrected drive frequencies, which should also be dominated by $\tilde{B}_{z0}$, to construct $\tilde{\epsilon_z}$ according to Eq.~\ref{ezdef} as
\be \label{ezcalc}
\tilde{\epsilon}_z = \tan^{-1}\left({1\over B_w}{\tilde{\upomega}_d^a+\tilde{\upomega}^b_d \over \gamma^a+\gamma^b}\right).
\ee
In the sections that follow we demonstrate the field suppression and stability of the comagnetometer. We will show that $\tilde{\epsilon}_z$ has a profound impact on the field suppression.

\begin{figure}
\includegraphics[width = 3.5in]{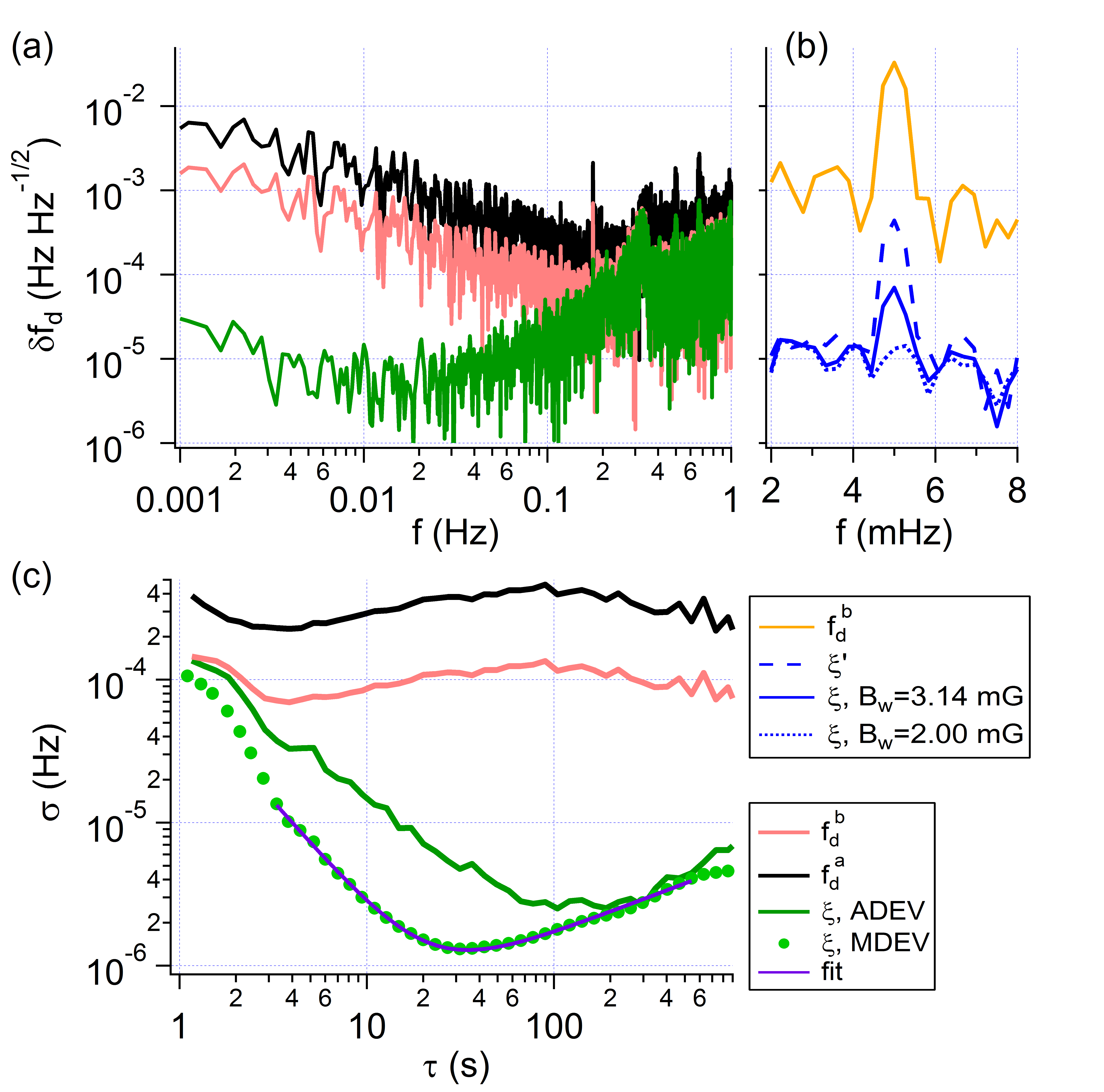}
\caption{ (a) Amplitude spectral densities of $f_d^{b}=\upomega_d^b/2\pi$, $f_d^{a}$, $\xi$. (b) Spectra with ancillary AC $B_z$ applied showing the field suppression factor for various comagnetometer signal computations. (c) Allan Deviations of $f_d^{b}$, $f_d^{a}$, $\xi$, where $\xi$ was analyzed with both standard and modified Allan deviations. All data shown have been corrected for finite gain. A Hanning window is applied to each spectrum.}\label{adev}
\end{figure}


\subsection{Field suppression} \label{FSF}

Each isotope is phase locked to line center with an accuracy of $\pm0.3$ mHz. The drive frequency of each isotope then tracks the resonance as defined by the phase of precession as measured by the Rb. These drive frequencies are recorded and corrected for finite gain (Eq.~\ref{upomeg}) in post-processing. The quantities $\tilde{\epsilon}_z$ and $\tilde{\xi}$ are also computed in post-processing (Eq.~\ref{xiprime} and ~\ref{ezcalc}).

Figure~\ref{adev} (a) shows the amplitude spectral densities of the drive frequencies and $\xi$. The spectrum of the $^{129}$Xe drive frequency is very nearly $\rho$ larger than the $^{131}$Xe drive frequency spectrum from DC to 0.1 Hz, suggesting that $\tilde{B}_{z0}$ dominates over this frequency band.

Since the goal of a comagnetometer is to sense non-magnetic spin-dependent interactions, we characterize how well the device rejects bias magnetic field perturbations. In order to characterize the field suppression factor (FSF) of the comagnetometer, we record the drive frequencies while applying an ancillary 5 mHz 4.3 $\mu$G $B_z$. We define the FSF to be $\tilde{\upomega}^b_d$/$\tilde{\xi}$ at the frequency of the ancillary $B_z$.

If we compute $\xi$ using the average of two independent measurements of $B_w$ (see next section) we find an FSF of 470. However, the maximum value of FSF occurs for $B_w=2.0$ mG and is 2300. Ignoring the influence of $\epsilon_z$ and computing the FSF as $\tilde{\upomega}^b_d$/$\tilde{\xi}'$ produces an FSF of merely 75. Clearly, $\epsilon_z$ introduces substantial phase shifts to the comagnetometer signal. In the following section we describe two independent measurements of $B_w$. We also describe a means of measuring the FSF without ambiguity introduced by uncertainty in $B_w$.

\subsubsection{$B_w$ measurements} \label{Bwmeas}

\begin{figure}
\includegraphics[width = 3.5in]{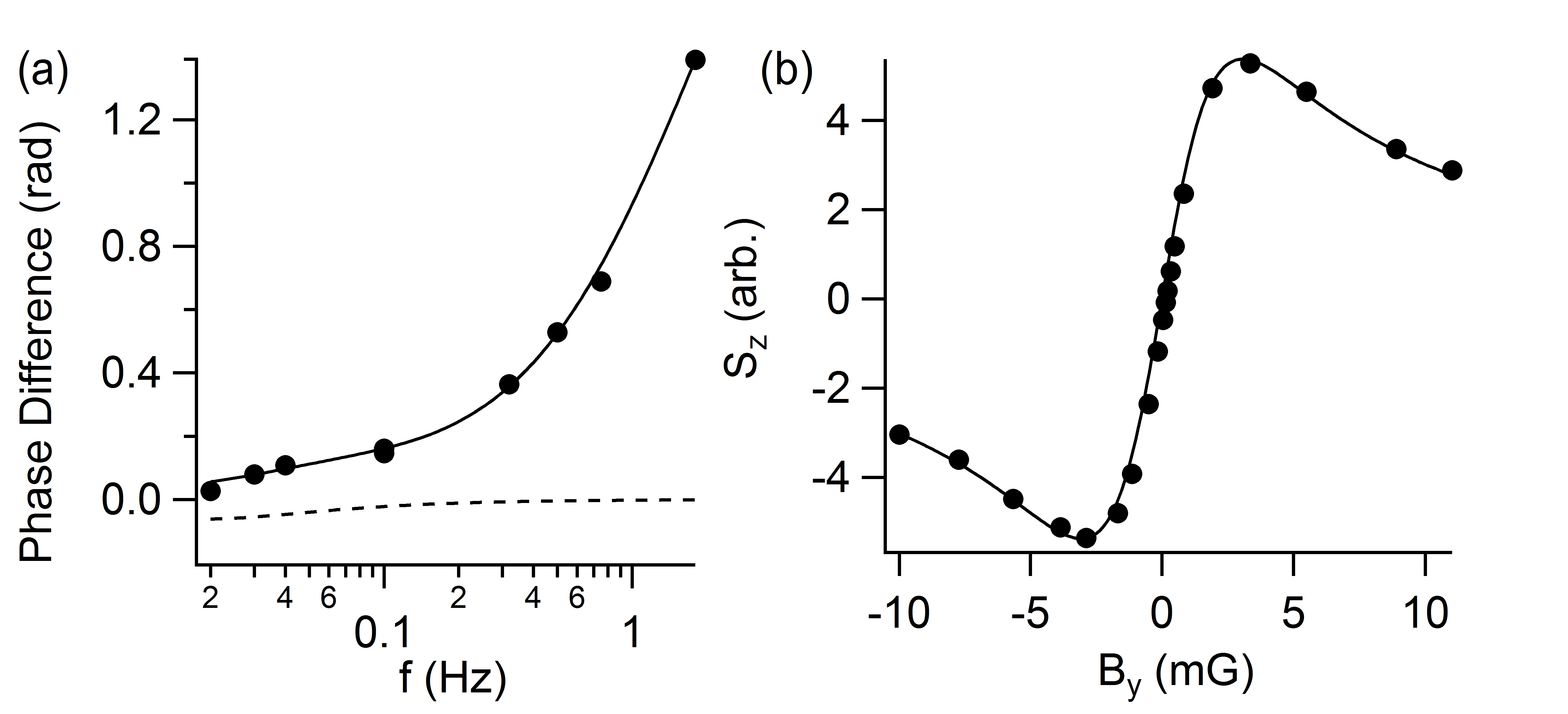}
\caption{(a) Open loop Xe phase difference. Dashed line is the calculated phase difference for $\tilde{\epsilon}_z=0$. Line is a fit that includes $\epsilon_z$. (b) Magnetometer response to $B_y$. Circles are measured data. Line is a dispersive fit to the data. }\label{Bws}
\end{figure}

We measure $B_w$ independently using two methods. First, we measure the difference in open loop response (see Eq.~\ref{delt2}) of the two Xe isotopes, which for $\omega>>\Gamma_2$ and $\Delta^a=\Delta^b=0$ is dominated by $\epsilon_z$. The data in Fig.~\ref{Bws} (a) were aquired by setting $\omega_d$ to resonance for each isotope by hand (no $\omega_d$ feedback), applying an AC few $\mu$G $B_{z0}$ at various frequencies, and recording the phase difference between the phase response of the two isotopes. Since we measure $\tilde{\delta} \mp \tilde{\epsilon}_z$ (as per Eq.~\ref{demodph}), we construct a fit function
\begin{multline}\label{phdiff}
\text{Arg}(\tilde{\delta}^a-\tilde{\epsilon}_z)-\text{Arg}(\tilde{\delta}^b+\tilde{\epsilon}_z)\\
\approx\text{tan}^{-1}\left({\omega(B_w \gamma^a-\Gamma_2^a)\over -\omega^2-B_w \gamma^a \Gamma_2^a}\right)-\text{tan}^{-1}\left({\omega(B_w \gamma^b+\Gamma_2^b)\over \omega^2-B_w \gamma^b \Gamma_2^b}\right),
\end{multline}
where we have taken $\tilde{\Delta}\approx \gamma \tilde{B}_{z0}$ and $\tilde{\epsilon}_z\approx \tilde{B}_{z0}/B_w$. Using this fit function, we find $B_w=3.5 \pm 0.3$ mG. If there were no magnetometer phase shift (i.e., if $\tilde{\epsilon}_z=0$) the phase difference would approach zero with increasing frequency (see dashed line in Fig.~\ref{Bws}). Instead we see the phase difference increase, consistent with our model for $\epsilon_z$.

We can also obtain $B_w$ by measuring the magnetometer response (without substantial Xe excitation) to a known $B_y$ of various magnitudes, as shown in Fig.~\ref{Bws} (b). Fitting according to the $\hat{z}$ component of Eq.~\ref{BE2}, we find $B_w = 3.1 \pm 0.1$ mG. The weighted mean of the two measurements is $B_w= 3.14$ mG. 

\subsubsection{Stabilizing the sum frequency}

\begin{figure}
\includegraphics[width=3.5in]{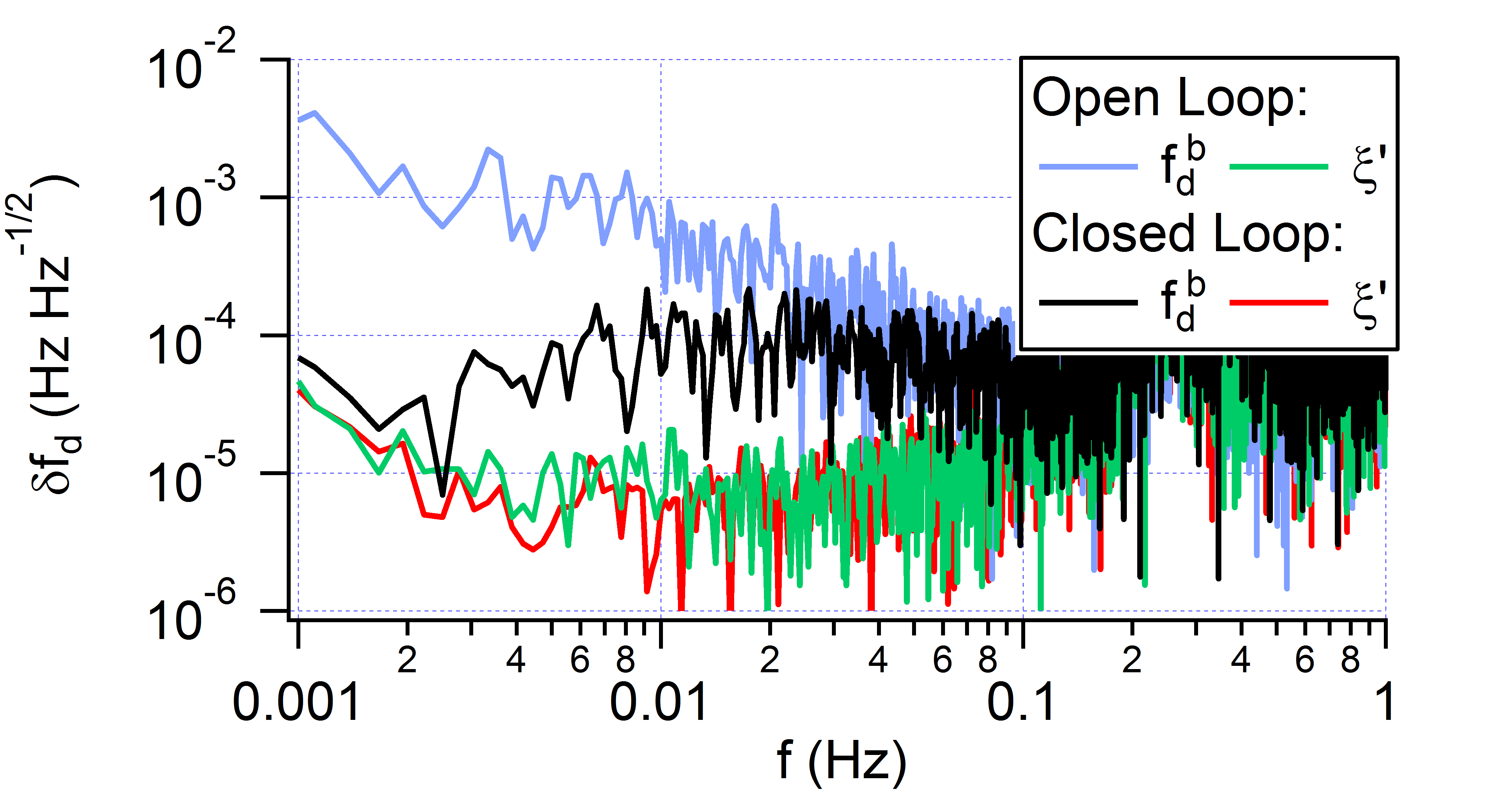}
\caption{Influence of stabilizing $\omega^a_d+\omega_d^b$ on $\tilde{f}^b_d$ and $\tilde{\xi}'$. Open loop data are without stabilization. Closed loop data are with stabilization.}\label{feedback}
\end{figure}

In order to prevent uncertainty in $B_w$ from influencing the FSF, we built an additional feedback loop which corrected $B_{z0}$ such that the sum $\omega_d^a+\omega_d^b$ was kept fixed. Assuming that $B_{z0}$ is the dominant contribution to $\epsilon_z$, this will stabilize the magnetometer phase shift. The gain of this feedback loop was sufficient to suppress $\tilde{\upomega}_d^b$ by a factor of 15 at 5 mHz compared to when the sum $\omega_d^a+\omega_d^b$ was not stabilized, as shown in Fig.~\ref{feedback}. The FSF was 1800 with this additional feedback loop (taking into acount the factor of 15 suppression of $\tilde{\upomega}_d^b$).

The low-frequency FSF of other comagnetometer implimentations are often not reported in the literature. The K-$^{21}$Ne comagnetometer from Ref.~\cite{Kornack2005} reports an FSF starting at $10^3$ at 1 Hz and decreasing linearly with applied $B_z$ frequency. Extrapolating to 5 mHz suggests an FSF of $\approx 10^4$ at 5 mHz, an order of magnitude greater than that which we demonstrate in this paper.

In the next section we demonstrate the stability of our comagnetometer. In particular we show that the stability is not limited by residual bias field fluctuations and/or finite FSF.

\subsection{Stability}

Without an ancillary $B_z$, $\xi$'s amplitude spectral density (see Fig.~\ref{adev}) has a distinct $f$ trending noise from 1 to 0.02 Hz until it reaches a white noise level of $<10$ $\mu$Hz /$\sqrt{\text{{Hz}}}$. Both the $f$ trending and white frequency noises are due to finite SNR and are discussed below. We find that the noise of $\xi$ is the same regardless of the $B_w$ used in computation and whether or not we stabilize $B_{z0}$ using the sum of the drive frequencies (both $\tilde{\xi}'$ in Fig.~\ref{feedback} are similar). This suggests that $\xi$'s noise is not limited by the FSF. A $B_w$ of $3.14$ mG was used for the $\xi$ shown in Fig.~\ref{adev} (a) and (c). 

Figure 6 (c) shows the standard Allan deviation (ADEV) of the drive frequencies and the comagnetometer signal $\xi$. It also depicts the modified Allan deviation (MDEV)~\cite{Allan1981,Vanier1989} of $\xi$. The ADEV of the drive frequencies are mostly constant with $\tau$ suggesting that they are dominated by flicker frequency noise. The high level of correlation between the two drive frequencies allows for a $\xi$ whose deviation is orders of magnitude smaller. The deviation of $\xi$ is dominated by white phase noise from 1 to 30 seconds of integration (as indicated by the $\tau^{-3/2}$ slope of the MDEV) and random walk of frequency from 30 seconds onward. 

We fit the MDEV from Fig.~\ref{adev} (c) using the function $\sigma^2 =(a\tau^{-3/2})^2+(b\tau^{-1/2})^2+(c\tau^{1/2})^2$ and find $a=78.2(7)$ $\mu$Hz Hz$^{-3/2}$, $b=4.3(4)$ $\mu$Hz Hz$^{-1/2}$, and $c=168(3)$ nHz Hz$^{1/2}$. The angle-random walk (ARW) is $b\sqrt{2}$ or $6.1(4)$ $\mu$Hz/$\sqrt{\text{Hz}}$.



We can estimate how finite SNR influences the comagnetometer stability by re-writing the comagnetometer signal to explicitly include the noise $\tilde{n}$ (SNR$^{-1}$) of our detection such that the measured phase of each isotope is $\tilde{\delta} \mp \tilde{\epsilon}_z+\tilde{n}$. The expression for $\tilde{\xi}^{\prime}$ (see Eq.~\ref{xiprime}) including the uncorrelated noise $\tilde{n}$ for each noble gas species becomes
\be
\tilde{\xi}^{\prime} = \tilde{\xi}+i\omega \tilde{\epsilon}_z+ {\rho (i\omega+\Gamma_2^b)\tilde{n}^b - (i\omega+\Gamma_2^a)\tilde{n}^a\over 1+\rho}.\label{mynoise}
\ee
If $\tilde{\xi}$ is dominated by $\tilde{n}$ then $\tilde{\xi}$ will have a white noise spectrum which changes to $f$ trending noise near $\omega\approx\Gamma_2$.

The white phase noise level measured under open loop conditions suggest SNRs of 5300 $\sqrt{\text{Hz}}$ and 3200 $\sqrt{\text{Hz}}$ for $^{131}$Xe and $^{129}$Xe, respectively. The SNR-limited ARW calculated from Eq.~\ref{mynoise} is 4 $\mu$Hz/$\sqrt{\text{Hz}}$, which is very similar to the ARW found by fitting the MDEV of $\xi$. The photon shot noise limited ARW is 10 nHz/$\sqrt{Hz}$. It is unclear what is presently limiting the SNR. We have confirmed that our detection noise is not limited by probe technical noise. It may be that our demodulation routine introduces noise. We may also be limited by finite bias pulse repetition rate modulation fidelity (drive-to-noise ratio). These are topics for future study.

\begin{figure}
\includegraphics[width = 2.8in]{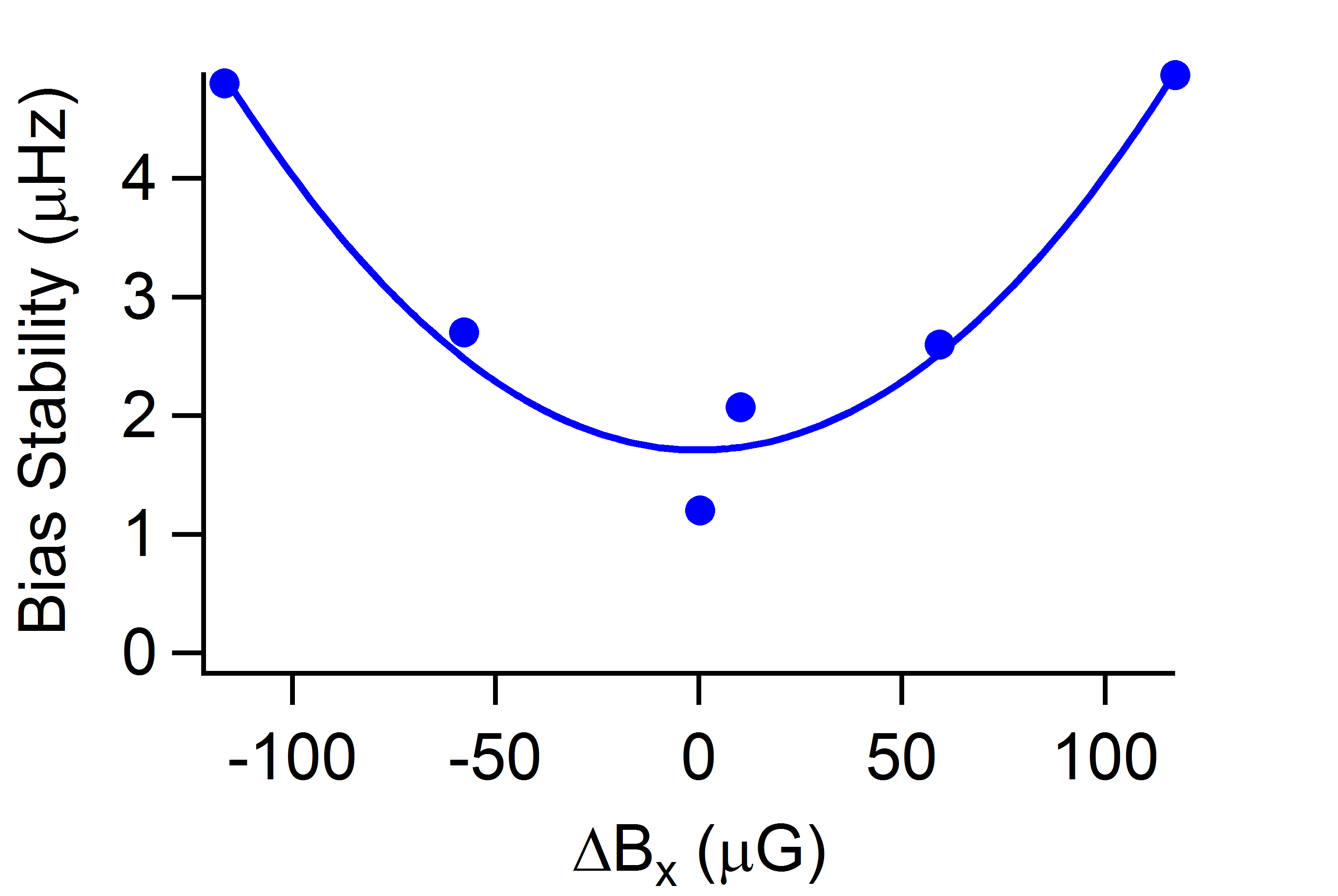}
\caption{Bias instability dependence on DC $B_x$. Line is a quadratic fit to the data.}\label{xcomp}
\end{figure}

It is uncertain what limits the bias instability. We find that the measured bias instability is most sensitive to a DC $B_x$, as shown in Fig.~\ref{xcomp}, which suggests that $1/f$ $B_x$ field fluctuations my be an important contributor to bias instability. We believe the mechanism for this dependence is related to imperfect cancellation of the Rb SE field, which depends critically on pump laser detuning, Rb density, and DC $B_x$. When the Rb SE field is not well cancelled, $K_z$ (a known source of systematic error) will be produced if $\Delta \neq 0$. We plan to study this effect in the near future. 

\section{Outlook} \label{outlook}
The accuracy of the device can be quantified by making measurements of a known non-magnetic spin-dependent phenomenon, such as Earth's rotation rate. This can be accomplished by orienting the sensitive axis of the comagnetometer parallel to Earth's axis of rotation. The demonstrated comagnetometer performance should allow resolution of Earth's rotation rate after less than 10 seconds of integration.

Further increase in comagnetometer sense bandwidth (beyond the 1 Hz bandwidth demonstrated here) is desirable for inertial navigation applications. The fundamental sense bandwidth of the comagnetometer is limited by $\omega_2$, which in this work was $\approx3.7$ Hz. The value of $\omega_2$ can be increased by increasing the average pulse repetition rate of the bias pulses or increasing their area to produce $2\pi m$ Larmor rotation of the Rb, where $m$ is an integer greater than one. However, in order to sample the $^{129}$Xe precession without violating Nyquist's criterion, the gating frequency $\omega_3$ (see Eq.~\ref{wpeq}) must be at least twice that of the average $^{129}$Xe precession frequency. The gating frequency is limited by the Rb magnetometer response. The magnetometer response can be increased via feedback by applying AC $B_y$ fields to cancel the Xe SE fields experienced by the Rb so that the Rb is kept at zero field. 

The performance of synchronous SE comagnetometry applied to $^3$He-$^{129}$Xe cells looks very promising. Compared to  $^{131}$Xe-$^{129}$Xe, a  $^3$He-$^{129}$Xe cell should enable roughly 10 times the $T_2$ and 10 times the signal~\cite{Walker2017,Limes2018}. If the modest magnetic sensitivity remained the same as our current cell then the anticipated ARW would be 100 nHz/$\sqrt{\text{Hz}}$, an order of magnitude less than what was reported in Ref.~\cite{Limes2018}. Although the improvement in SNR should be dramatic, the sensitivity to longitudinal Rb fields will be much larger ($b_S^{a}  = 110 b_S^{\text{He}}$\cite{Ma2011} versus $b_S^{a}  = 1.002 b_S^{b}$\cite{Bulatowicz2013,Petrov2019}), as will the sensitivity to first-order temperature gradients~\cite{Sheng2014} and back polarization~\cite{Limes2018}.

The through-space J-coupling term $b_a^b$ depends on cell geometry and the frequency enhancement factor $\kappa_a^b$ which has yet to be measured but was recently calculated to be 0.3~\cite{Vaara2019,Limes2019}. We estimate that for our apparatus $K_z^{a} = 0.1K_{max}^{a}$ would produce a comagnetometer frequency shift of a few $\mu$Hz, which would be well resolved by this device.

Dual-species synchronous SEOP is an excellent technique for performing a direct search for axion-induced spin-mass couplings. Because of the Yukawa type potentials assumed for spin-mass couplings, miniaturized comagnetometers enable broad energy resolution of the supposed coupling. If the results reported in this study are reproduced in a 2 mm cell~\cite{Bulatowicz2013}, we anticipate being able to improve the present upper bound in the submillimeter wavelength range by an order of magnitude.  

\section{Conclusion} \label{conclusion}

We have demonstrated the first SE pumped comagnetometer which continuously polarizes alkali-metal atoms and noble gas nuclei in a purely transverse manner. The pure transverse polarization suppresses the influence of $K_z$ and $S_z$ on the comagnetometer signal. The application of the bias field as a sequence of low duty cycle pulses whose area correspond to $2\pi$ precession of the alkali-metal atoms enables magnetic field modulations which are experienced by the noble gas and not by the alkali-metal atoms.

The embedded alkali-metal magnetometer was used to detect the precession phase of two noble gas isotopes simultaneously with a sample rate of 200 Hz. The measured $^{129}$Xe polarization appears to be limited by a temperature-dependent wall relaxation mechanism. This effect is independent of the continuous excitation method described in this work and likely stems from the cell's Rb hydride coating. The comagnetometer was operated in a closed loop fashion such that the drive frequencies of each isotope were corrected to keep their measured phases equal to zero. The closed loop noble gas precession frequencies were shown to be sufficiently well correlated to enable a field suppression factor of $10^3$, which was more than sufficient to suppress the low frequency field noise such that the white noise of the device was apparent. Our system has a 1 Hz measurement bandwidth, and the ARW was found to be 7 $\mu$Hz /$\sqrt{\text{Hz}}$, within a factor of two of the SNR limit. To first order, the comagnetometer scale factor can be written in terms of fundamental constants. 

We have demonstrated dramatic improvement of the field suppression factor when accounting for the phase of the alkali-metal magnetometer. This is accomplished by computing $\epsilon_z$ from the stray field noise in $B_z$ (which we determine from the sum of the recorded Xe drive frequencies) and including it in the computation of the comagnetometer signal. A similar field suppression factor was achieved by stabilizing the stray field in real time. While our calculations assumed $\epsilon_z$ was dominated by $B_z$, other possible contributions include $K_z$, pump pointing, and back polarization ($K_y$ producing $S_y$). We could monitor $\epsilon_z$ directly by applying an ancillary rotating $B_{\perp}$ and measuring the response of the magnetometer, allowing for improved suppression of the magnetometer phase shift in the comagnetometry.

\begin{acknowledgements}

We acknowledge Anna Korver and Josh Weber for prior supporting work, and Michael Bulatowicz for his help in designing the pulse driver circuit. We would like to thank Michael Larsen for insightful discussions. This research was supported by the National Science Foundation (PHY-1607439 and PHY-1912543) and Northrop Grumman Mission Systems' University Research Program.

\end{acknowledgements}

\bibliography{spinexchangecopy}

\begin{thebibliography}{43}%
\makeatletter
\providecommand \@ifxundefined [1]{%
 \@ifx{#1\undefined}
}%
\providecommand \@ifnum [1]{%
 \ifnum #1\expandafter \@firstoftwo
 \else \expandafter \@secondoftwo
 \fi
}%
\providecommand \@ifx [1]{%
 \ifx #1\expandafter \@firstoftwo
 \else \expandafter \@secondoftwo
 \fi
}%
\providecommand \natexlab [1]{#1}%
\providecommand \enquote  [1]{``#1''}%
\providecommand \bibnamefont  [1]{#1}%
\providecommand \bibfnamefont [1]{#1}%
\providecommand \citenamefont [1]{#1}%
\providecommand \href@noop [0]{\@secondoftwo}%
\providecommand \href [0]{\begingroup \@sanitize@url \@href}%
\providecommand \@href[1]{\@@startlink{#1}\@@href}%
\providecommand \@@href[1]{\endgroup#1\@@endlink}%
\providecommand \@sanitize@url [0]{\catcode `\\12\catcode `\$12\catcode
  `\&12\catcode `\#12\catcode `\^12\catcode `\_12\catcode `\%12\relax}%
\providecommand \@@startlink[1]{}%
\providecommand \@@endlink[0]{}%
\providecommand \url  [0]{\begingroup\@sanitize@url \@url }%
\providecommand \@url [1]{\endgroup\@href {#1}{\urlprefix }}%
\providecommand \urlprefix  [0]{URL }%
\providecommand \Eprint [0]{\href }%
\providecommand \doibase [0]{http://dx.doi.org/}%
\providecommand \selectlanguage [0]{\@gobble}%
\providecommand \bibinfo  [0]{\@secondoftwo}%
\providecommand \bibfield  [0]{\@secondoftwo}%
\providecommand \translation [1]{[#1]}%
\providecommand \BibitemOpen [0]{}%
\providecommand \bibitemStop [0]{}%
\providecommand \bibitemNoStop [0]{.\EOS\space}%
\providecommand \EOS [0]{\spacefactor3000\relax}%
\providecommand \BibitemShut  [1]{\csname bibitem#1\endcsname}%
\let\auto@bib@innerbib\@empty
\bibitem [{\citenamefont {Limes}\ \emph {et~al.}(2018)\citenamefont {Limes},
  \citenamefont {Sheng},\ and\ \citenamefont {Romalis}}]{Limes2018}%
  \BibitemOpen
  \bibfield  {author} {\bibinfo {author} {\bibfnamefont {M.~E.}\ \bibnamefont
  {Limes}}, \bibinfo {author} {\bibfnamefont {D.}~\bibnamefont {Sheng}}, \ and\
  \bibinfo {author} {\bibfnamefont {M.~V.}\ \bibnamefont {Romalis}},\
  }\bibfield  {title} {\enquote {\bibinfo {title}
  {$^{3}\mathrm{He}\text{\ensuremath{-}}^{129}\mathrm{Xe}$ comagnetometery
  using $^{87}\mathrm{Rb}$ detection and decoupling},}\ }\href {\doibase
  10.1103/PhysRevLett.120.033401} {\bibfield  {journal} {\bibinfo  {journal}
  {Phys. Rev. Lett.}\ }\textbf {\bibinfo {volume} {120}},\ \bibinfo {pages}
  {033401} (\bibinfo {year} {2018})}\BibitemShut {NoStop}%
\bibitem [{\citenamefont {Walker}\ and\ \citenamefont
  {Larsen}(2016)}]{Walker2016}%
  \BibitemOpen
  \bibfield  {author} {\bibinfo {author} {\bibfnamefont {T.G.}\ \bibnamefont
  {Walker}}\ and\ \bibinfo {author} {\bibfnamefont {M.S.}\ \bibnamefont
  {Larsen}},\ }\bibfield  {title} {\enquote {\bibinfo {title}
  {Spin-exchange-pumped nmr gyros},}\ }\href {\doibase
  https://doi.org/10.1016/bs.aamop.2016.04.002} {\ \bibinfo {series} {Advances
  In Atomic, Molecular, and Optical Physics},\ \textbf {\bibinfo {volume}
  {65}},\ \bibinfo {pages} {373 -- 401} (\bibinfo {year} {2016})}\BibitemShut
  {NoStop}%
\bibitem [{\citenamefont {Kornack}\ \emph {et~al.}(2005)\citenamefont
  {Kornack}, \citenamefont {Ghosh},\ and\ \citenamefont
  {Romalis}}]{Kornack2005}%
  \BibitemOpen
  \bibfield  {author} {\bibinfo {author} {\bibfnamefont {T.~W.}\ \bibnamefont
  {Kornack}}, \bibinfo {author} {\bibfnamefont {R.~K.}\ \bibnamefont {Ghosh}},
  \ and\ \bibinfo {author} {\bibfnamefont {M.~V.}\ \bibnamefont {Romalis}},\
  }\bibfield  {title} {\enquote {\bibinfo {title} {{Nuclear spin gyroscope
  based on an atomic comagnetometer}},}\ }\href@noop {} {\bibfield  {journal}
  {\bibinfo  {journal} {Physical review letters}\ }\textbf {\bibinfo {volume}
  {95}},\ \bibinfo {pages} {230801} (\bibinfo {year} {2005})}\BibitemShut
  {NoStop}%
\bibitem [{\citenamefont {Jiang}\ \emph {et~al.}(2018)\citenamefont {Jiang},
  \citenamefont {Quan}, \citenamefont {Li}, \citenamefont {Fan}, \citenamefont
  {Liu}, \citenamefont {Qin}, \citenamefont {Wan},\ and\ \citenamefont
  {Fang}}]{Jiang2018}%
  \BibitemOpen
  \bibfield  {author} {\bibinfo {author} {\bibfnamefont {Liwei}\ \bibnamefont
  {Jiang}}, \bibinfo {author} {\bibfnamefont {Wei}\ \bibnamefont {Quan}},
  \bibinfo {author} {\bibfnamefont {Rujie}\ \bibnamefont {Li}}, \bibinfo
  {author} {\bibfnamefont {Wenfeng}\ \bibnamefont {Fan}}, \bibinfo {author}
  {\bibfnamefont {Feng}\ \bibnamefont {Liu}}, \bibinfo {author} {\bibfnamefont
  {Jie}\ \bibnamefont {Qin}}, \bibinfo {author} {\bibfnamefont {Shuangai}\
  \bibnamefont {Wan}}, \ and\ \bibinfo {author} {\bibfnamefont {Jiancheng}\
  \bibnamefont {Fang}},\ }\bibfield  {title} {\enquote {\bibinfo {title} {A
  parametrically modulated dual-axis atomic spin gyroscope},}\ }\href {\doibase
  10.1063/1.5018015} {\bibfield  {journal} {\bibinfo  {journal} {Applied
  Physics Letters}\ }\textbf {\bibinfo {volume} {112}},\ \bibinfo {pages}
  {054103} (\bibinfo {year} {2018})},\ \Eprint
  {http://arxiv.org/abs/https://doi.org/10.1063/1.5018015}
  {https://doi.org/10.1063/1.5018015} \BibitemShut {NoStop}%
\bibitem [{\citenamefont {Karwacki}(1980)}]{Karwacki1980}%
  \BibitemOpen
  \bibfield  {author} {\bibinfo {author} {\bibfnamefont {F.~A.}\ \bibnamefont
  {Karwacki}},\ }\bibfield  {title} {\enquote {\bibinfo {title} {Nuclear
  magnetic resonance gyro development},}\ }\href@noop {} {\bibfield  {journal}
  {\bibinfo  {journal} {Navigation}\ }\textbf {\bibinfo {volume} {27}},\
  \bibinfo {pages} {72} (\bibinfo {year} {1980})}\BibitemShut {NoStop}%
\bibitem [{\citenamefont {Bulatowicz}\ \emph {et~al.}(2013)\citenamefont
  {Bulatowicz}, \citenamefont {Griffith}, \citenamefont {Larsen}, \citenamefont
  {Mirijanian}, \citenamefont {Fu}, \citenamefont {Smith}, \citenamefont
  {Snow}, \citenamefont {Yan},\ and\ \citenamefont {Walker}}]{Bulatowicz2013}%
  \BibitemOpen
  \bibfield  {author} {\bibinfo {author} {\bibfnamefont {M.}~\bibnamefont
  {Bulatowicz}}, \bibinfo {author} {\bibfnamefont {R.}~\bibnamefont
  {Griffith}}, \bibinfo {author} {\bibfnamefont {M.}~\bibnamefont {Larsen}},
  \bibinfo {author} {\bibfnamefont {J.}~\bibnamefont {Mirijanian}}, \bibinfo
  {author} {\bibfnamefont {C.~B.}\ \bibnamefont {Fu}}, \bibinfo {author}
  {\bibfnamefont {E.}~\bibnamefont {Smith}}, \bibinfo {author} {\bibfnamefont
  {W.~M.}\ \bibnamefont {Snow}}, \bibinfo {author} {\bibfnamefont
  {H.}~\bibnamefont {Yan}}, \ and\ \bibinfo {author} {\bibfnamefont {T.~G.}\
  \bibnamefont {Walker}},\ }\bibfield  {title} {\enquote {\bibinfo {title}
  {Laboratory search for a long-range ${T}$-odd, ${P}$-odd interaction from
  axionlike particles using dual-species nuclear magnetic resonance with
  polarized $^{129}\mathrm{Xe}$ and $^{131}\mathrm{Xe}$ gas},}\ }\href
  {\doibase 10.1103/PhysRevLett.111.102001} {\bibfield  {journal} {\bibinfo
  {journal} {Phys. Rev. Lett.}\ }\textbf {\bibinfo {volume} {111}},\ \bibinfo
  {pages} {102001} (\bibinfo {year} {2013})}\BibitemShut {NoStop}%
\bibitem [{\citenamefont {Lee}\ \emph {et~al.}(2018)\citenamefont {Lee},
  \citenamefont {Almasi},\ and\ \citenamefont {Romalis}}]{Lee2018}%
  \BibitemOpen
  \bibfield  {author} {\bibinfo {author} {\bibfnamefont {J.}~\bibnamefont
  {Lee}}, \bibinfo {author} {\bibfnamefont {A.}~\bibnamefont {Almasi}}, \ and\
  \bibinfo {author} {\bibfnamefont {M.~V.}\ \bibnamefont {Romalis}},\
  }\bibfield  {title} {\enquote {\bibinfo {title} {Improved limits on spin-mass
  interactions},}\ }\href {\doibase 10.1103/PhysRevLett.120.161801} {\bibfield
  {journal} {\bibinfo  {journal} {Phys. Rev. Lett.}\ }\textbf {\bibinfo
  {volume} {120}},\ \bibinfo {pages} {161801} (\bibinfo {year}
  {2018})}\BibitemShut {NoStop}%
\bibitem [{\citenamefont {Allmendinger}\ \emph {et~al.}(2014)\citenamefont
  {Allmendinger}, \citenamefont {Heil}, \citenamefont {Karpuk}, \citenamefont
  {Kilian}, \citenamefont {Scharth}, \citenamefont {Schmidt}, \citenamefont
  {Schnabel}, \citenamefont {Sobolev},\ and\ \citenamefont
  {Tullney}}]{Allmendinger2014}%
  \BibitemOpen
  \bibfield  {author} {\bibinfo {author} {\bibfnamefont {F.}~\bibnamefont
  {Allmendinger}}, \bibinfo {author} {\bibfnamefont {W.}~\bibnamefont {Heil}},
  \bibinfo {author} {\bibfnamefont {S.}~\bibnamefont {Karpuk}}, \bibinfo
  {author} {\bibfnamefont {W.}~\bibnamefont {Kilian}}, \bibinfo {author}
  {\bibfnamefont {A.}~\bibnamefont {Scharth}}, \bibinfo {author} {\bibfnamefont
  {U.}~\bibnamefont {Schmidt}}, \bibinfo {author} {\bibfnamefont
  {A.}~\bibnamefont {Schnabel}}, \bibinfo {author} {\bibfnamefont {Yu.}\
  \bibnamefont {Sobolev}}, \ and\ \bibinfo {author} {\bibfnamefont
  {K.}~\bibnamefont {Tullney}},\ }\bibfield  {title} {\enquote {\bibinfo
  {title} {New limit on lorentz-invariance- and ${CPT}$-violating neutron spin
  interactions using a free-spin-precession $^{3}${He}-$^{129}${Xe}
  comagnetometer},}\ }\href {\doibase 10.1103/PhysRevLett.112.110801}
  {\bibfield  {journal} {\bibinfo  {journal} {Phys. Rev. Lett.}\ }\textbf
  {\bibinfo {volume} {112}},\ \bibinfo {pages} {110801} (\bibinfo {year}
  {2014})}\BibitemShut {NoStop}%
\bibitem [{\citenamefont {Romalis}\ \emph {et~al.}(2014)\citenamefont
  {Romalis}, \citenamefont {Sheng}, \citenamefont {Saam},\ and\ \citenamefont
  {Walker}}]{Romalis2014}%
  \BibitemOpen
  \bibfield  {author} {\bibinfo {author} {\bibfnamefont {M.~V.}\ \bibnamefont
  {Romalis}}, \bibinfo {author} {\bibfnamefont {D.}~\bibnamefont {Sheng}},
  \bibinfo {author} {\bibfnamefont {B.}~\bibnamefont {Saam}}, \ and\ \bibinfo
  {author} {\bibfnamefont {T.~G.}\ \bibnamefont {Walker}},\ }\bibfield  {title}
  {\enquote {\bibinfo {title} {Comment on ``new limit on lorentz-invariance-
  and ${CPT}$-violating neutron spin interactions using a free-spin-precession
  $^{3}${He}-$^{129}${Xe} comagnetometer''},}\ }\href {\doibase
  10.1103/PhysRevLett.113.188901} {\bibfield  {journal} {\bibinfo  {journal}
  {Phys. Rev. Lett.}\ }\textbf {\bibinfo {volume} {113}},\ \bibinfo {pages}
  {188901} (\bibinfo {year} {2014})}\BibitemShut {NoStop}%
\bibitem [{\citenamefont {Smiciklas}\ \emph {et~al.}(2011)\citenamefont
  {Smiciklas}, \citenamefont {Brown}, \citenamefont {Cheuk}, \citenamefont
  {Smullin},\ and\ \citenamefont {Romalis}}]{Smiciklas2011}%
  \BibitemOpen
  \bibfield  {author} {\bibinfo {author} {\bibfnamefont {M.}~\bibnamefont
  {Smiciklas}}, \bibinfo {author} {\bibfnamefont {J.~M.}\ \bibnamefont
  {Brown}}, \bibinfo {author} {\bibfnamefont {L.~W.}\ \bibnamefont {Cheuk}},
  \bibinfo {author} {\bibfnamefont {S.~J.}\ \bibnamefont {Smullin}}, \ and\
  \bibinfo {author} {\bibfnamefont {M.~V.}\ \bibnamefont {Romalis}},\
  }\bibfield  {title} {\enquote {\bibinfo {title} {New test of local lorentz
  invariance using a
  $^{21}\mathrm{Ne}\mathrm{\text{-}}\mathrm{Rb}\mathrm{\text{-}}\mathrm{K}$
  comagnetometer},}\ }\href {\doibase 10.1103/PhysRevLett.107.171604}
  {\bibfield  {journal} {\bibinfo  {journal} {Phys. Rev. Lett.}\ }\textbf
  {\bibinfo {volume} {107}},\ \bibinfo {pages} {171604} (\bibinfo {year}
  {2011})}\BibitemShut {NoStop}%
\bibitem [{\citenamefont {Brown}\ \emph {et~al.}(2010)\citenamefont {Brown},
  \citenamefont {Smullin}, \citenamefont {Kornack},\ and\ \citenamefont
  {Romalis}}]{Brown2010}%
  \BibitemOpen
  \bibfield  {author} {\bibinfo {author} {\bibfnamefont {J.~M.}\ \bibnamefont
  {Brown}}, \bibinfo {author} {\bibfnamefont {S.~J.}\ \bibnamefont {Smullin}},
  \bibinfo {author} {\bibfnamefont {T.~W.}\ \bibnamefont {Kornack}}, \ and\
  \bibinfo {author} {\bibfnamefont {M.~V.}\ \bibnamefont {Romalis}},\
  }\bibfield  {title} {\enquote {\bibinfo {title} {New limit on lorentz- and
  ${CPT}$-violating neutron spin interactions},}\ }\href {\doibase
  10.1103/PhysRevLett.105.151604} {\bibfield  {journal} {\bibinfo  {journal}
  {Phys. Rev. Lett.}\ }\textbf {\bibinfo {volume} {105}},\ \bibinfo {pages}
  {151604} (\bibinfo {year} {2010})}\BibitemShut {NoStop}%
\bibitem [{\citenamefont {Glenday}\ \emph {et~al.}(2008)\citenamefont
  {Glenday}, \citenamefont {Cramer}, \citenamefont {Phillips},\ and\
  \citenamefont {Walsworth}}]{Glenday2008}%
  \BibitemOpen
  \bibfield  {author} {\bibinfo {author} {\bibfnamefont {A.~G.}\ \bibnamefont
  {Glenday}}, \bibinfo {author} {\bibfnamefont {C.~E.}\ \bibnamefont {Cramer}},
  \bibinfo {author} {\bibfnamefont {D.~F.}\ \bibnamefont {Phillips}}, \ and\
  \bibinfo {author} {\bibfnamefont {R.~L.}\ \bibnamefont {Walsworth}},\
  }\bibfield  {title} {\enquote {\bibinfo {title} {Limits on anomalous
  spin-spin couplings between neutrons},}\ }\href {\doibase
  10.1103/PhysRevLett.101.261801} {\bibfield  {journal} {\bibinfo  {journal}
  {Phys. Rev. Lett.}\ }\textbf {\bibinfo {volume} {101}},\ \bibinfo {pages}
  {261801} (\bibinfo {year} {2008})}\BibitemShut {NoStop}%
\bibitem [{\citenamefont {Rosenberry}\ and\ \citenamefont
  {Chupp}(2001)}]{Rosenberry2001}%
  \BibitemOpen
  \bibfield  {author} {\bibinfo {author} {\bibfnamefont {M.~A.}\ \bibnamefont
  {Rosenberry}}\ and\ \bibinfo {author} {\bibfnamefont {T.~E.}\ \bibnamefont
  {Chupp}},\ }\bibfield  {title} {\enquote {\bibinfo {title} {Atomic electric
  dipole moment measurement using spin exchange pumped masers of
  ${}^{129}\mathrm{Xe}$ and ${}^{3}\mathrm{He}$},}\ }\href {\doibase
  10.1103/PhysRevLett.86.22} {\bibfield  {journal} {\bibinfo  {journal} {Phys.
  Rev. Lett.}\ }\textbf {\bibinfo {volume} {86}},\ \bibinfo {pages} {22--25}
  (\bibinfo {year} {2001})}\BibitemShut {NoStop}%
\bibitem [{\citenamefont {Allmendinger}\ \emph {et~al.}(2019)\citenamefont
  {Allmendinger}, \citenamefont {Engin}, \citenamefont {Heil}, \citenamefont
  {Karpuk}, \citenamefont {Krause}, \citenamefont {Niederl\"ander},
  \citenamefont {Offenh\"ausser}, \citenamefont {Repetto}, \citenamefont
  {Schmidt},\ and\ \citenamefont {Zimmer}}]{Allmendinger2019}%
  \BibitemOpen
  \bibfield  {author} {\bibinfo {author} {\bibfnamefont {F.}~\bibnamefont
  {Allmendinger}}, \bibinfo {author} {\bibfnamefont {I.}~\bibnamefont {Engin}},
  \bibinfo {author} {\bibfnamefont {W.}~\bibnamefont {Heil}}, \bibinfo {author}
  {\bibfnamefont {S.}~\bibnamefont {Karpuk}}, \bibinfo {author} {\bibfnamefont
  {H.-J.}\ \bibnamefont {Krause}}, \bibinfo {author} {\bibfnamefont
  {B.}~\bibnamefont {Niederl\"ander}}, \bibinfo {author} {\bibfnamefont
  {A.}~\bibnamefont {Offenh\"ausser}}, \bibinfo {author} {\bibfnamefont
  {M.}~\bibnamefont {Repetto}}, \bibinfo {author} {\bibfnamefont
  {U.}~\bibnamefont {Schmidt}}, \ and\ \bibinfo {author} {\bibfnamefont
  {S.}~\bibnamefont {Zimmer}},\ }\bibfield  {title} {\enquote {\bibinfo {title}
  {Measurement of the permanent electric dipole moment of the
  $^{129}\mathrm{Xe}$ atom},}\ }\href {\doibase 10.1103/PhysRevA.100.022505}
  {\bibfield  {journal} {\bibinfo  {journal} {Phys. Rev. A}\ }\textbf {\bibinfo
  {volume} {100}},\ \bibinfo {pages} {022505} (\bibinfo {year}
  {2019})}\BibitemShut {NoStop}%
\bibitem [{\citenamefont {Sachdeva}\ \emph {et~al.}(2019)\citenamefont
  {Sachdeva}, \citenamefont {Fan}, \citenamefont {Babcock}, \citenamefont
  {Burghoff}, \citenamefont {Chupp}, \citenamefont {Degenkolb}, \citenamefont
  {Fierlinger}, \citenamefont {Haude}, \citenamefont {Kraegeloh}, \citenamefont
  {Kilian}, \citenamefont {Knappe-Gr\"uneberg}, \citenamefont {Kuchler},
  \citenamefont {Liu}, \citenamefont {Marino}, \citenamefont {Meinel},
  \citenamefont {Rolfs}, \citenamefont {Salhi}, \citenamefont {Schnabel},
  \citenamefont {Singh}, \citenamefont {Stuiber}, \citenamefont {Terrano},
  \citenamefont {Trahms},\ and\ \citenamefont {Voigt}}]{Sachdeva2019}%
  \BibitemOpen
  \bibfield  {author} {\bibinfo {author} {\bibfnamefont {N.}~\bibnamefont
  {Sachdeva}}, \bibinfo {author} {\bibfnamefont {I.}~\bibnamefont {Fan}},
  \bibinfo {author} {\bibfnamefont {E.}~\bibnamefont {Babcock}}, \bibinfo
  {author} {\bibfnamefont {M.}~\bibnamefont {Burghoff}}, \bibinfo {author}
  {\bibfnamefont {T.~E.}\ \bibnamefont {Chupp}}, \bibinfo {author}
  {\bibfnamefont {S.}~\bibnamefont {Degenkolb}}, \bibinfo {author}
  {\bibfnamefont {P.}~\bibnamefont {Fierlinger}}, \bibinfo {author}
  {\bibfnamefont {S.}~\bibnamefont {Haude}}, \bibinfo {author} {\bibfnamefont
  {E.}~\bibnamefont {Kraegeloh}}, \bibinfo {author} {\bibfnamefont
  {W.}~\bibnamefont {Kilian}}, \bibinfo {author} {\bibfnamefont
  {S.}~\bibnamefont {Knappe-Gr\"uneberg}}, \bibinfo {author} {\bibfnamefont
  {F.}~\bibnamefont {Kuchler}}, \bibinfo {author} {\bibfnamefont
  {T.}~\bibnamefont {Liu}}, \bibinfo {author} {\bibfnamefont {M.}~\bibnamefont
  {Marino}}, \bibinfo {author} {\bibfnamefont {J.}~\bibnamefont {Meinel}},
  \bibinfo {author} {\bibfnamefont {K.}~\bibnamefont {Rolfs}}, \bibinfo
  {author} {\bibfnamefont {Z.}~\bibnamefont {Salhi}}, \bibinfo {author}
  {\bibfnamefont {A.}~\bibnamefont {Schnabel}}, \bibinfo {author}
  {\bibfnamefont {J.~T.}\ \bibnamefont {Singh}}, \bibinfo {author}
  {\bibfnamefont {S.}~\bibnamefont {Stuiber}}, \bibinfo {author} {\bibfnamefont
  {W.~A.}\ \bibnamefont {Terrano}}, \bibinfo {author} {\bibfnamefont
  {L.}~\bibnamefont {Trahms}}, \ and\ \bibinfo {author} {\bibfnamefont
  {J.}~\bibnamefont {Voigt}},\ }\bibfield  {title} {\enquote {\bibinfo {title}
  {New limit on the permanent electric dipole moment of $^{129}\mathrm{Xe}$
  using $^{3}\mathrm{He}$ comagnetometry and squid detection},}\ }\href
  {\doibase 10.1103/PhysRevLett.123.143003} {\bibfield  {journal} {\bibinfo
  {journal} {Phys. Rev. Lett.}\ }\textbf {\bibinfo {volume} {123}},\ \bibinfo
  {pages} {143003} (\bibinfo {year} {2019})}\BibitemShut {NoStop}%
\bibitem [{\citenamefont {Volk}\ \emph {et~al.}({1980})\citenamefont {Volk},
  \citenamefont {Kwon},\ and\ \citenamefont {Mark}}]{Volk80b}%
  \BibitemOpen
  \bibfield  {author} {\bibinfo {author} {\bibfnamefont {C.~H.}\ \bibnamefont
  {Volk}}, \bibinfo {author} {\bibfnamefont {T.M.}\ \bibnamefont {Kwon}}, \
  and\ \bibinfo {author} {\bibfnamefont {J.~G.}\ \bibnamefont {Mark}},\
  }\bibfield  {title} {\enquote {\bibinfo {title} {Measurement of the
  $^{87}\mathrm{Rb}-^{129}\mathrm{Xe}$ spin-exchange cross section},}\ }\href
  {\doibase {10.1103/PhysRevA.21.1549}} {\bibfield  {journal} {\bibinfo
  {journal} {Phys. Rev. A}\ }\textbf {\bibinfo {volume} {{21}}},\ \bibinfo
  {pages} {{1549--1555}} (\bibinfo {year} {{1980}})}\BibitemShut {NoStop}%
\bibitem [{\citenamefont {Kwon}\ \emph {et~al.}(1981)\citenamefont {Kwon},
  \citenamefont {Mark},\ and\ \citenamefont {Volk}}]{Kwon1981}%
  \BibitemOpen
  \bibfield  {author} {\bibinfo {author} {\bibfnamefont {T.~M.}\ \bibnamefont
  {Kwon}}, \bibinfo {author} {\bibfnamefont {J.~G.}\ \bibnamefont {Mark}}, \
  and\ \bibinfo {author} {\bibfnamefont {C.~H.}\ \bibnamefont {Volk}},\
  }\bibfield  {title} {\enquote {\bibinfo {title} {Quadrupole nuclear spin
  relaxation of $^{131}\mathrm{Xe}$ in the presence of rubidium vapor},}\
  }\href {\doibase 10.1103/PhysRevA.24.1894} {\bibfield  {journal} {\bibinfo
  {journal} {Phys. Rev. A}\ }\textbf {\bibinfo {volume} {24}},\ \bibinfo
  {pages} {1894--1903} (\bibinfo {year} {1981})}\BibitemShut {NoStop}%
\bibitem [{\citenamefont {Donley}(2010)}]{Donley2010}%
  \BibitemOpen
  \bibfield  {author} {\bibinfo {author} {\bibfnamefont {E.A.}\ \bibnamefont
  {Donley}},\ }\bibfield  {title} {\enquote {\bibinfo {title} {Nuclear magnetic
  resonance gyroscopes},}\ }in\ \href {\doibase 10.1109/ICSENS.2010.5690983}
  {\emph {\bibinfo {booktitle} {Sensors, 2010 IEEE}}}\ (\bibinfo {year}
  {2010})\ pp.\ \bibinfo {pages} {17 --22}\BibitemShut {NoStop}%
\bibitem [{\citenamefont {Walker}\ and\ \citenamefont
  {Happer}(1997)}]{Walker1997}%
  \BibitemOpen
  \bibfield  {author} {\bibinfo {author} {\bibfnamefont {T.G.}\ \bibnamefont
  {Walker}}\ and\ \bibinfo {author} {\bibfnamefont {W.}~\bibnamefont
  {Happer}},\ }\bibfield  {title} {\enquote {\bibinfo {title} {{Spin-exchange
  optical pumping of noble-gas nuclei}},}\ }\href
  {http://link.aps.org/doi/10.1103/RevModPhys.69.629} {\bibfield  {journal}
  {\bibinfo  {journal} {Rev. Mod. Phys.}\ }\textbf {\bibinfo {volume} {69}},\
  \bibinfo {pages} {629--642} (\bibinfo {year} {1997})}\BibitemShut {NoStop}%
\bibitem [{\citenamefont {Limes}\ \emph {et~al.}(2019)\citenamefont {Limes},
  \citenamefont {Dural}, \citenamefont {Romalis}, \citenamefont {Foley},
  \citenamefont {Kornack}, \citenamefont {Nelson}, \citenamefont {Grisham},\
  and\ \citenamefont {Vaara}}]{Limes2019}%
  \BibitemOpen
  \bibfield  {author} {\bibinfo {author} {\bibfnamefont {M.~E.}\ \bibnamefont
  {Limes}}, \bibinfo {author} {\bibfnamefont {N.}~\bibnamefont {Dural}},
  \bibinfo {author} {\bibfnamefont {M.~V.}\ \bibnamefont {Romalis}}, \bibinfo
  {author} {\bibfnamefont {E.~L.}\ \bibnamefont {Foley}}, \bibinfo {author}
  {\bibfnamefont {T.~W.}\ \bibnamefont {Kornack}}, \bibinfo {author}
  {\bibfnamefont {A.}~\bibnamefont {Nelson}}, \bibinfo {author} {\bibfnamefont
  {L.~R.}\ \bibnamefont {Grisham}}, \ and\ \bibinfo {author} {\bibfnamefont
  {J.}~\bibnamefont {Vaara}},\ }\bibfield  {title} {\enquote {\bibinfo {title}
  {Dipolar and scalar $^{3}\mathrm{He}\text{\ensuremath{-}}^{129}\mathrm{Xe}$
  frequency shifts in stemless cells},}\ }\href {\doibase
  10.1103/PhysRevA.100.010501} {\bibfield  {journal} {\bibinfo  {journal}
  {Phys. Rev. A}\ }\textbf {\bibinfo {volume} {100}},\ \bibinfo {pages}
  {010501(R)} (\bibinfo {year} {2019})}\BibitemShut {NoStop}%
\bibitem [{\citenamefont {Terrano}\ \emph {et~al.}(2019)\citenamefont
  {Terrano}, \citenamefont {Meinel}, \citenamefont {Sachdeva}, \citenamefont
  {Chupp}, \citenamefont {Degenkolb}, \citenamefont {Fierlinger}, \citenamefont
  {Kuchler},\ and\ \citenamefont {Singh}}]{Terrano2019}%
  \BibitemOpen
  \bibfield  {author} {\bibinfo {author} {\bibfnamefont {W.~A.}\ \bibnamefont
  {Terrano}}, \bibinfo {author} {\bibfnamefont {J.}~\bibnamefont {Meinel}},
  \bibinfo {author} {\bibfnamefont {N.}~\bibnamefont {Sachdeva}}, \bibinfo
  {author} {\bibfnamefont {T.~E.}\ \bibnamefont {Chupp}}, \bibinfo {author}
  {\bibfnamefont {S.}~\bibnamefont {Degenkolb}}, \bibinfo {author}
  {\bibfnamefont {P.}~\bibnamefont {Fierlinger}}, \bibinfo {author}
  {\bibfnamefont {F.}~\bibnamefont {Kuchler}}, \ and\ \bibinfo {author}
  {\bibfnamefont {J.~T.}\ \bibnamefont {Singh}},\ }\bibfield  {title} {\enquote
  {\bibinfo {title} {Frequency shifts in noble-gas comagnetometers},}\ }\href
  {\doibase 10.1103/PhysRevA.100.012502} {\bibfield  {journal} {\bibinfo
  {journal} {Phys. Rev. A}\ }\textbf {\bibinfo {volume} {100}},\ \bibinfo
  {pages} {012502} (\bibinfo {year} {2019})}\BibitemShut {NoStop}%
\bibitem [{\citenamefont {{Petrov}}\ \emph {et~al.}(2019)\citenamefont
  {{Petrov}}, \citenamefont {{Pazgalev}},\ and\ \citenamefont
  {{Vershovskii}}}]{Petrov2019}%
  \BibitemOpen
  \bibfield  {author} {\bibinfo {author} {\bibfnamefont {V.~I.}\ \bibnamefont
  {{Petrov}}}, \bibinfo {author} {\bibfnamefont {A.~S.}\ \bibnamefont
  {{Pazgalev}}}, \ and\ \bibinfo {author} {\bibfnamefont {A.~K.}\ \bibnamefont
  {{Vershovskii}}},\ }\bibfield  {title} {\enquote {\bibinfo {title} {Isotope
  shift of nuclear magnetic resonances in 129{Xe} and 131{Xe} caused by
  spin-exchange pumping by alkali metal atoms},}\ }\href {\doibase
  10.1109/JSEN.2019.2945002} {\bibfield  {journal} {\bibinfo  {journal} {IEEE
  Sensors Journal}\ ,\ \bibinfo {pages} {1--1}} (\bibinfo {year}
  {2019})}\BibitemShut {NoStop}%
\bibitem [{\citenamefont {Schaefer}\ \emph {et~al.}(1989)\citenamefont
  {Schaefer}, \citenamefont {Cates}, \citenamefont {Chien}, \citenamefont
  {Gonatas}, \citenamefont {Happer},\ and\ \citenamefont
  {Walker}}]{Schaefer89}%
  \BibitemOpen
  \bibfield  {author} {\bibinfo {author} {\bibfnamefont {S.~R.}\ \bibnamefont
  {Schaefer}}, \bibinfo {author} {\bibfnamefont {G.~D.}\ \bibnamefont {Cates}},
  \bibinfo {author} {\bibfnamefont {Ting-Ray}\ \bibnamefont {Chien}}, \bibinfo
  {author} {\bibfnamefont {D.}~\bibnamefont {Gonatas}}, \bibinfo {author}
  {\bibfnamefont {W.}~\bibnamefont {Happer}}, \ and\ \bibinfo {author}
  {\bibfnamefont {T.~G.}\ \bibnamefont {Walker}},\ }\bibfield  {title}
  {\enquote {\bibinfo {title} {Frequency shifts of the magnetic-resonance
  spectrum of mixtures of nuclear spin-polarized noble gases and vapors of
  spin-polarized alkali-metal atoms},}\ }\href {\doibase
  10.1103/PhysRevA.39.5613} {\bibfield  {journal} {\bibinfo  {journal} {Phys.
  Rev. A}\ }\textbf {\bibinfo {volume} {39}},\ \bibinfo {pages} {5613--5623}
  (\bibinfo {year} {1989})}\BibitemShut {NoStop}%
\bibitem [{\citenamefont {Chupp}\ \emph {et~al.}(1988)\citenamefont {Chupp},
  \citenamefont {Oteiza}, \citenamefont {Richardson},\ and\ \citenamefont
  {White}}]{Chupp1988}%
  \BibitemOpen
  \bibfield  {author} {\bibinfo {author} {\bibfnamefont {T.~E.}\ \bibnamefont
  {Chupp}}, \bibinfo {author} {\bibfnamefont {E.~R.}\ \bibnamefont {Oteiza}},
  \bibinfo {author} {\bibfnamefont {J.~M.}\ \bibnamefont {Richardson}}, \ and\
  \bibinfo {author} {\bibfnamefont {T.~R.}\ \bibnamefont {White}},\ }\bibfield
  {title} {\enquote {\bibinfo {title} {Precision frequency measurements with
  polarized $^{3}\mathrm{He}$, $^{21}\mathrm{Ne}$, and $^{129}\mathrm{Xe}$
  atoms},}\ }\href {\doibase 10.1103/PhysRevA.38.3998} {\bibfield  {journal}
  {\bibinfo  {journal} {Phys. Rev. A}\ }\textbf {\bibinfo {volume} {38}},\
  \bibinfo {pages} {3998--4003} (\bibinfo {year} {1988})}\BibitemShut {NoStop}%
\bibitem [{\citenamefont {Takahashi}\ \emph {et~al.}(1999)\citenamefont
  {Takahashi}, \citenamefont {Honda}, \citenamefont {Tanaka}, \citenamefont
  {Toyoda}, \citenamefont {Ishikawa},\ and\ \citenamefont
  {Yabuzaki}}]{Takahashi1999}%
  \BibitemOpen
  \bibfield  {author} {\bibinfo {author} {\bibfnamefont {Y.}~\bibnamefont
  {Takahashi}}, \bibinfo {author} {\bibfnamefont {K.}~\bibnamefont {Honda}},
  \bibinfo {author} {\bibfnamefont {N.}~\bibnamefont {Tanaka}}, \bibinfo
  {author} {\bibfnamefont {K.}~\bibnamefont {Toyoda}}, \bibinfo {author}
  {\bibfnamefont {K.}~\bibnamefont {Ishikawa}}, \ and\ \bibinfo {author}
  {\bibfnamefont {T.}~\bibnamefont {Yabuzaki}},\ }\bibfield  {title} {\enquote
  {\bibinfo {title} {Quantum nondemolition measurement of spin via the
  paramagnetic faraday rotation},}\ }\href {\doibase 10.1103/PhysRevA.60.4974}
  {\bibfield  {journal} {\bibinfo  {journal} {Phys. Rev. A}\ }\textbf {\bibinfo
  {volume} {60}},\ \bibinfo {pages} {4974--4979} (\bibinfo {year}
  {1999})}\BibitemShut {NoStop}%
\bibitem [{\citenamefont {Katz}\ \emph {et~al.}(2019)\citenamefont {Katz},
  \citenamefont {Shaham},\ and\ \citenamefont {Firstenberg}}]{Katz2019}%
  \BibitemOpen
  \bibfield  {author} {\bibinfo {author} {\bibfnamefont {O.}~\bibnamefont
  {Katz}}, \bibinfo {author} {\bibfnamefont {R.}~\bibnamefont {Shaham}}, \ and\
  \bibinfo {author} {\bibfnamefont {O.}~\bibnamefont {Firstenberg}},\
  }\href@noop {} {\enquote {\bibinfo {title} {Quantum interface for noble-gas
  spins},}\ } (\bibinfo {year} {2019}),\ \Eprint
  {http://arxiv.org/abs/1905.12532} {arXiv:1905.12532 [quant-ph]} \BibitemShut
  {NoStop}%
\bibitem [{\citenamefont {Nahlawi}\ \emph {et~al.}(2019)\citenamefont
  {Nahlawi}, \citenamefont {Ma}, \citenamefont {Conradi},\ and\ \citenamefont
  {Saam}}]{Nahlawi2019}%
  \BibitemOpen
  \bibfield  {author} {\bibinfo {author} {\bibfnamefont {A.~I.}\ \bibnamefont
  {Nahlawi}}, \bibinfo {author} {\bibfnamefont {Z.~L.}\ \bibnamefont {Ma}},
  \bibinfo {author} {\bibfnamefont {M.~S.}\ \bibnamefont {Conradi}}, \ and\
  \bibinfo {author} {\bibfnamefont {B.}~\bibnamefont {Saam}},\ }\bibfield
  {title} {\enquote {\bibinfo {title} {High-precision determination of the
  frequency-shift enhancement factor in $\mathrm{Rb}-^{129}\mathrm{Xe}$},}\
  }\href {\doibase 10.1103/PhysRevA.100.053415} {\bibfield  {journal} {\bibinfo
   {journal} {Phys. Rev. A}\ }\textbf {\bibinfo {volume} {100}},\ \bibinfo
  {pages} {053415} (\bibinfo {year} {2019})}\BibitemShut {NoStop}%
\bibitem [{\citenamefont {Makulski}(2015)}]{Makulski2015}%
  \BibitemOpen
  \bibfield  {author} {\bibinfo {author} {\bibfnamefont {W.}~\bibnamefont
  {Makulski}},\ }\bibfield  {title} {\enquote {\bibinfo {title} {$^{129}${Xe}
  and $^{131}${Xe} nuclear magnetic dipole moments from gas phase nmr
  spectra},}\ }\href {\doibase 10.1002/mrc.4191} {\bibfield  {journal}
  {\bibinfo  {journal} {Magnetic Resonance in Chemistry}\ }\textbf {\bibinfo
  {volume} {53}},\ \bibinfo {pages} {273--279} (\bibinfo {year} {2015})},\
  \Eprint
  {http://arxiv.org/abs/https://onlinelibrary.wiley.com/doi/pdf/10.1002/mrc.4191}
  {https://onlinelibrary.wiley.com/doi/pdf/10.1002/mrc.4191} \BibitemShut
  {NoStop}%
\bibitem [{\citenamefont {Korver}\ \emph {et~al.}(2015)\citenamefont {Korver},
  \citenamefont {Thrasher}, \citenamefont {Bulatowicz},\ and\ \citenamefont
  {Walker}}]{Korver2015}%
  \BibitemOpen
  \bibfield  {author} {\bibinfo {author} {\bibfnamefont {A.}~\bibnamefont
  {Korver}}, \bibinfo {author} {\bibfnamefont {D.}~\bibnamefont {Thrasher}},
  \bibinfo {author} {\bibfnamefont {M.}~\bibnamefont {Bulatowicz}}, \ and\
  \bibinfo {author} {\bibfnamefont {T.~G.}\ \bibnamefont {Walker}},\ }\bibfield
   {title} {\enquote {\bibinfo {title} {Synchronous spin-exchange optical
  pumping},}\ }\href {\doibase 10.1103/PhysRevLett.115.253001} {\bibfield
  {journal} {\bibinfo  {journal} {Phys. Rev. Lett.}\ }\textbf {\bibinfo
  {volume} {115}},\ \bibinfo {pages} {253001} (\bibinfo {year}
  {2015})}\BibitemShut {NoStop}%
\bibitem [{\citenamefont {Thrasher}\ \emph {et~al.}(2019)\citenamefont
  {Thrasher}, \citenamefont {Sorensen}, \citenamefont {Weber}, \citenamefont
  {Bulatowicz}, \citenamefont {Korver}, \citenamefont {Larsen},\ and\
  \citenamefont {Walker}}]{Thrasher2019}%
  \BibitemOpen
  \bibfield  {author} {\bibinfo {author} {\bibfnamefont {D.~A.}\ \bibnamefont
  {Thrasher}}, \bibinfo {author} {\bibfnamefont {S.~S.}\ \bibnamefont
  {Sorensen}}, \bibinfo {author} {\bibfnamefont {J.}~\bibnamefont {Weber}},
  \bibinfo {author} {\bibfnamefont {M.}~\bibnamefont {Bulatowicz}}, \bibinfo
  {author} {\bibfnamefont {A.}~\bibnamefont {Korver}}, \bibinfo {author}
  {\bibfnamefont {M.}~\bibnamefont {Larsen}}, \ and\ \bibinfo {author}
  {\bibfnamefont {T.~G.}\ \bibnamefont {Walker}},\ }\bibfield  {title}
  {\enquote {\bibinfo {title} {Continuous comagnetometry using transversely
  polarized xe isotopes},}\ }\href {\doibase 10.1103/PhysRevA.100.061403}
  {\bibfield  {journal} {\bibinfo  {journal} {Phys. Rev. A}\ }\textbf {\bibinfo
  {volume} {100}},\ \bibinfo {pages} {061403(R)} (\bibinfo {year}
  {2019})}\BibitemShut {NoStop}%
\bibitem [{\citenamefont {Brinkmann}\ \emph {et~al.}(1962)\citenamefont
  {Brinkmann}, \citenamefont {Brun},\ and\ \citenamefont
  {Staub}}]{Brinkmann62}%
  \BibitemOpen
  \bibfield  {author} {\bibinfo {author} {\bibfnamefont {D.}~\bibnamefont
  {Brinkmann}}, \bibinfo {author} {\bibfnamefont {E.}~\bibnamefont {Brun}}, \
  and\ \bibinfo {author} {\bibfnamefont {H.~H.}\ \bibnamefont {Staub}},\
  }\bibfield  {title} {\enquote {\bibinfo {title} {Kernresonanz im gasformigen
  xenon},}\ }\href@noop {} {\bibfield  {journal} {\bibinfo  {journal} {Helv.
  Phys. Acta}\ }\textbf {\bibinfo {volume} {35}},\ \bibinfo {pages} {431}
  (\bibinfo {year} {1962})}\BibitemShut {NoStop}%
\bibitem [{\citenamefont {Korver}\ \emph {et~al.}(2013)\citenamefont {Korver},
  \citenamefont {Wyllie}, \citenamefont {Lancor},\ and\ \citenamefont
  {Walker}}]{Korver2013}%
  \BibitemOpen
  \bibfield  {author} {\bibinfo {author} {\bibfnamefont {A.}~\bibnamefont
  {Korver}}, \bibinfo {author} {\bibfnamefont {R.}~\bibnamefont {Wyllie}},
  \bibinfo {author} {\bibfnamefont {B.}~\bibnamefont {Lancor}}, \ and\ \bibinfo
  {author} {\bibfnamefont {T.~G.}\ \bibnamefont {Walker}},\ }\bibfield  {title}
  {\enquote {\bibinfo {title} {Suppression of spin-exchange relaxation using
  pulsed parametric resonance},}\ }\href {\doibase
  10.1103/PhysRevLett.111.043002} {\bibfield  {journal} {\bibinfo  {journal}
  {Phys. Rev. Lett.}\ }\textbf {\bibinfo {volume} {111}},\ \bibinfo {pages}
  {043002} (\bibinfo {year} {2013})}\BibitemShut {NoStop}%
\bibitem [{\citenamefont {Bhaskar}\ \emph {et~al.}(1980)\citenamefont
  {Bhaskar}, \citenamefont {Hou}, \citenamefont {Ligare}, \citenamefont
  {Suleman},\ and\ \citenamefont {Happer}}]{Bhaskar80b}%
  \BibitemOpen
  \bibfield  {author} {\bibinfo {author} {\bibfnamefont {N.~D.}\ \bibnamefont
  {Bhaskar}}, \bibinfo {author} {\bibfnamefont {M.}~\bibnamefont {Hou}},
  \bibinfo {author} {\bibfnamefont {M.}~\bibnamefont {Ligare}}, \bibinfo
  {author} {\bibfnamefont {B.}~\bibnamefont {Suleman}}, \ and\ \bibinfo
  {author} {\bibfnamefont {W.}~\bibnamefont {Happer}},\ }\bibfield  {title}
  {\enquote {\bibinfo {title} {Role of {Na}-{Xe} molecules in spin relaxation
  of optically pumped {Na} in {Xe} gas},}\ }\href {\doibase
  10.1103/PhysRevA.22.2710} {\bibfield  {journal} {\bibinfo  {journal} {Phys.
  Rev. A}\ }\textbf {\bibinfo {volume} {22}},\ \bibinfo {pages} {2710--2716}
  (\bibinfo {year} {1980})}\BibitemShut {NoStop}%
\bibitem [{\citenamefont {Babcock}\ \emph {et~al.}(2006)\citenamefont
  {Babcock}, \citenamefont {Chann}, \citenamefont {Walker}, \citenamefont
  {Chen},\ and\ \citenamefont {Gentile}}]{Babcock2006}%
  \BibitemOpen
  \bibfield  {author} {\bibinfo {author} {\bibfnamefont {E.}~\bibnamefont
  {Babcock}}, \bibinfo {author} {\bibfnamefont {B.}~\bibnamefont {Chann}},
  \bibinfo {author} {\bibfnamefont {T.~G.}\ \bibnamefont {Walker}}, \bibinfo
  {author} {\bibfnamefont {W.~C.}\ \bibnamefont {Chen}}, \ and\ \bibinfo
  {author} {\bibfnamefont {T.~R.}\ \bibnamefont {Gentile}},\ }\bibfield
  {title} {\enquote {\bibinfo {title} {Limits to the polarization for
  spin-exchange optical pumping of $^{3}\mathrm{He}$},}\ }\href {\doibase
  10.1103/PhysRevLett.96.083003} {\bibfield  {journal} {\bibinfo  {journal}
  {Phys. Rev. Lett.}\ }\textbf {\bibinfo {volume} {96}},\ \bibinfo {pages}
  {083003} (\bibinfo {year} {2006})}\BibitemShut {NoStop}%
\bibitem [{\citenamefont {Wu}\ \emph {et~al.}(1988)\citenamefont {Wu},
  \citenamefont {Schaefer}, \citenamefont {Cates},\ and\ \citenamefont
  {Happer}}]{Wu1988}%
  \BibitemOpen
  \bibfield  {author} {\bibinfo {author} {\bibfnamefont {Z.}~\bibnamefont
  {Wu}}, \bibinfo {author} {\bibfnamefont {S.}~\bibnamefont {Schaefer}},
  \bibinfo {author} {\bibfnamefont {G.~D.}\ \bibnamefont {Cates}}, \ and\
  \bibinfo {author} {\bibfnamefont {W.}~\bibnamefont {Happer}},\ }\bibfield
  {title} {\enquote {\bibinfo {title} {Coherent interactions of the polarized
  nuclear spins of gaseous atoms with the container walls},}\ }\href {\doibase
  10.1103/PhysRevA.37.1161} {\bibfield  {journal} {\bibinfo  {journal} {Phys.
  Rev. A}\ }\textbf {\bibinfo {volume} {37}},\ \bibinfo {pages} {1161--1175}
  (\bibinfo {year} {1988})}\BibitemShut {NoStop}%
\bibitem [{\citenamefont {Donley}\ \emph {et~al.}(2009)\citenamefont {Donley},
  \citenamefont {Long}, \citenamefont {Liebisch}, \citenamefont {Hodby},
  \citenamefont {Fisher},\ and\ \citenamefont {Kitching}}]{Donley2009}%
  \BibitemOpen
  \bibfield  {author} {\bibinfo {author} {\bibfnamefont {E.~A.}\ \bibnamefont
  {Donley}}, \bibinfo {author} {\bibfnamefont {J.~L.}\ \bibnamefont {Long}},
  \bibinfo {author} {\bibfnamefont {T.~C.}\ \bibnamefont {Liebisch}}, \bibinfo
  {author} {\bibfnamefont {E.~R.}\ \bibnamefont {Hodby}}, \bibinfo {author}
  {\bibfnamefont {T.~A.}\ \bibnamefont {Fisher}}, \ and\ \bibinfo {author}
  {\bibfnamefont {J.}~\bibnamefont {Kitching}},\ }\bibfield  {title} {\enquote
  {\bibinfo {title} {Nuclear quadrupole resonances in compact vapor cells: The
  crossover between the nmr and the nuclear quadrupole resonance interaction
  regimes},}\ }\href {\doibase 10.1103/PhysRevA.79.013420} {\bibfield
  {journal} {\bibinfo  {journal} {Phys. Rev. A}\ }\textbf {\bibinfo {volume}
  {79}},\ \bibinfo {pages} {013420} (\bibinfo {year} {2009})}\BibitemShut
  {NoStop}%
\bibitem [{\citenamefont {Bechhoefer}(2005)}]{Bechhoefer2005}%
  \BibitemOpen
  \bibfield  {author} {\bibinfo {author} {\bibfnamefont {J.}~\bibnamefont
  {Bechhoefer}},\ }\bibfield  {title} {\enquote {\bibinfo {title} {Feedback for
  physicists: A tutorial essay on control},}\ }\href {\doibase
  10.1103/RevModPhys.77.783} {\bibfield  {journal} {\bibinfo  {journal} {Rev.
  Mod. Phys.}\ }\textbf {\bibinfo {volume} {77}},\ \bibinfo {pages} {783--836}
  (\bibinfo {year} {2005})}\BibitemShut {NoStop}%
\bibitem [{\citenamefont {Allan}\ and\ \citenamefont {Barnes}()}]{Allan1981}%
  \BibitemOpen
  \bibfield  {author} {\bibinfo {author} {\bibfnamefont {D.W.}\ \bibnamefont
  {Allan}}\ and\ \bibinfo {author} {\bibfnamefont {J.A.}\ \bibnamefont
  {Barnes}},\ }\bibfield  {title} {\enquote {\bibinfo {title} {A modified
  "allan variance" with increased oscillator characterization ability},}\
  }\href@noop {} {\bibinfo  {journal} {Proceedings of the 35th Anual Frequency
  Control Symposium, Philadelphia, 1981}\ ,\ \bibinfo {pages}
  {470--475}}\BibitemShut {NoStop}%
\bibitem [{\citenamefont {Vanier}\ and\ \citenamefont
  {Audoin}(1989)}]{Vanier1989}%
  \BibitemOpen
\bibfield  {journal} {  }\bibfield  {author} {\bibinfo {author} {\bibfnamefont
  {J.}~\bibnamefont {Vanier}}\ and\ \bibinfo {author} {\bibfnamefont
  {C.}~\bibnamefont {Audoin}},\ }in\ \href@noop {} {\emph {\bibinfo {booktitle}
  {{The Quantum Physics of Atomic Frequency Standards}}}},\ Vol.~\bibinfo
  {volume} {1}\ (\bibinfo  {publisher} {IOP Publishing Ltd, Bristol},\ \bibinfo
  {year} {1989})\ Chap.\ \bibinfo {chapter} {Appendix 2F}, pp.\ \bibinfo
  {pages} {216--256}\BibitemShut {NoStop}%
\bibitem [{\citenamefont {Gentile}\ \emph {et~al.}(2017)\citenamefont
  {Gentile}, \citenamefont {Nacher}, \citenamefont {Saam},\ and\ \citenamefont
  {Walker}}]{Walker2017}%
  \BibitemOpen
  \bibfield  {author} {\bibinfo {author} {\bibfnamefont {T.~R.}\ \bibnamefont
  {Gentile}}, \bibinfo {author} {\bibfnamefont {P.~J.}\ \bibnamefont {Nacher}},
  \bibinfo {author} {\bibfnamefont {B.}~\bibnamefont {Saam}}, \ and\ \bibinfo
  {author} {\bibfnamefont {T.~G.}\ \bibnamefont {Walker}},\ }\bibfield  {title}
  {\enquote {\bibinfo {title} {Optically polarized $^{3}\mathrm{He}$},}\ }\href
  {\doibase 10.1103/RevModPhys.89.045004} {\bibfield  {journal} {\bibinfo
  {journal} {Rev. Mod. Phys.}\ }\textbf {\bibinfo {volume} {89}},\ \bibinfo
  {pages} {045004} (\bibinfo {year} {2017})}\BibitemShut {NoStop}%
\bibitem [{\citenamefont {Ma}\ \emph {et~al.}(2011)\citenamefont {Ma},
  \citenamefont {Sorte},\ and\ \citenamefont {Saam}}]{Ma2011}%
  \BibitemOpen
  \bibfield  {author} {\bibinfo {author} {\bibfnamefont {Z.~L.}\ \bibnamefont
  {Ma}}, \bibinfo {author} {\bibfnamefont {E.~G.}\ \bibnamefont {Sorte}}, \
  and\ \bibinfo {author} {\bibfnamefont {B.}~\bibnamefont {Saam}},\ }\bibfield
  {title} {\enquote {\bibinfo {title} {Collisional $^{3}\mathrm{He}$ and
  $^{129}\mathrm{Xe}$ frequency shifts in rb\char21{}noble-gas mixtures},}\
  }\href {\doibase 10.1103/PhysRevLett.106.193005} {\bibfield  {journal}
  {\bibinfo  {journal} {Phys. Rev. Lett.}\ }\textbf {\bibinfo {volume} {106}},\
  \bibinfo {pages} {193005} (\bibinfo {year} {2011})}\BibitemShut {NoStop}%
\bibitem [{\citenamefont {Sheng}\ \emph {et~al.}(2014)\citenamefont {Sheng},
  \citenamefont {Kabcenell},\ and\ \citenamefont {Romalis}}]{Sheng2014}%
  \BibitemOpen
  \bibfield  {author} {\bibinfo {author} {\bibfnamefont {D.}~\bibnamefont
  {Sheng}}, \bibinfo {author} {\bibfnamefont {A.}~\bibnamefont {Kabcenell}}, \
  and\ \bibinfo {author} {\bibfnamefont {M.~V.}\ \bibnamefont {Romalis}},\
  }\bibfield  {title} {\enquote {\bibinfo {title} {New classes of systematic
  effects in gas spin comagnetometers},}\ }\href {\doibase
  10.1103/PhysRevLett.113.163002} {\bibfield  {journal} {\bibinfo  {journal}
  {Phys. Rev. Lett.}\ }\textbf {\bibinfo {volume} {113}},\ \bibinfo {pages}
  {163002} (\bibinfo {year} {2014})}\BibitemShut {NoStop}%
\bibitem [{\citenamefont {Vaara}\ and\ \citenamefont
  {Romalis}(2019)}]{Vaara2019}%
  \BibitemOpen
  \bibfield  {author} {\bibinfo {author} {\bibfnamefont {J.}~\bibnamefont
  {Vaara}}\ and\ \bibinfo {author} {\bibfnamefont {M.~V.}\ \bibnamefont
  {Romalis}},\ }\bibfield  {title} {\enquote {\bibinfo {title} {Calculation of
  scalar nuclear spin-spin coupling in a noble-gas mixture},}\ }\href {\doibase
  10.1103/PhysRevA.99.060501} {\bibfield  {journal} {\bibinfo  {journal} {Phys.
  Rev. A}\ }\textbf {\bibinfo {volume} {99}},\ \bibinfo {pages} {060501(R)}
  (\bibinfo {year} {2019})}\BibitemShut {NoStop}%
\end{thebibliography}%

\end{document}